\documentclass[onecolumn,useAMS,usenatbib]{mn2e}

\usepackage{graphicx}
\usepackage{subfigure}

\usepackage{amssymb}

\def\beq{\begin{equation}}
\def\eeq{\end{equation}}
\def\rmd{{\rm d}}

\title[]{Radiation drag in the field of a non-spherical source}

\author[D. Bini et al.]
{D. Bini$^{1,2}$\thanks{E-mail:binid@icra.it},
A. Geralico$^{1}$\thanks{E-mail:geralico@icra.it}
,
A. Passamonti$^3$\thanks{E-mail:andrea.passamonti@oa-roma.inaf.it}
\\ \\
$^1$ Istituto per le Applicazioni del Calcolo ``M. Picone,'' CNR, I-00185 Rome, Italy\\
$^2$ INFN, Sezione di Napoli, I--80126 Naples, Italy\\
$^3$ INAF - Osservatorio Astronomico di Roma, via Frascati 44, I-00040, Monteporzio Catone (Roma), Italy
}

\begin{document}

\date{\today}

\pagerange{\pageref{firstpage}--\pageref{lastpage}} \pubyear{}

\maketitle

\label{firstpage}

\begin{abstract}
The motion of a test particle in the gravitational field of a non-spherical source endowed with both mass and mass quadrupole moment is investigated when a test radiation field is also present. 
The background is described by the Erez-Rosen solution, which is a static spacetime belonging to the Weyl class of solutions to the vacuum Einstein's field equations, and reduces to the familiar Schwarzschild solution when the quadrupole parameter vanishes. 
The radiation flux has a fixed but arbitrary (non-zero) angular momentum.
The interaction with the radiation field is assumed to be Thomson-like, i.e., the particles absorb and re-emit radiation, thus suffering for a friction-like drag force.
Such an additional force is responsible for the Poynting-Robertson effect, which is well established in the framework of Newtonian gravity and has been recently extended to the general theory of relativity.
The balance between gravitational attraction, centrifugal force and radiation drag leads to the occurrence of equilibrium circular orbits which are attractors for the surrounding matter for every fixed value of the interaction strength.
The presence of the quadrupolar structure of the source introduces a further degree of freedom: there exists a whole family of equilibrium orbits parametrized by the quadrupole parameter, generalizing previous works. 
This scenario is expected to play a role in the context of accretion matter around compact objects.
\end{abstract}


\section{Introduction} 

Most of the compact objects currently investigated and modeled in astrophysics are \lq\lq observable" because of the presence of some matter or radiation around them (we will often term them as \lq\lq central objects" below). 
At the lowest order of approximation, the associated gravitational field can be described by the spherically symmetric Schwarzschild solution, when shape deformation and rotation effects of the source can be neglected, and matter can be considered as a collection of test particles moving along geodesics.
However, this is only an idealized situation which is likely not to represent any real astrophysical system. 
The dynamics of a body orbiting a compact object is indeed strongly influenced by the combined effect of the overall gravitational field, the surrounding matter and in turn the radiation distribution associated with high energy processes arising thereby.
A typical example is represented by a luminous compact object surrounded by orbiting matter, such as accretion discs around a neutron star or a black hole. 
The most luminous and persistent sources of electromagnetic radiation in the universe are the active galactic nuclei (AGN), which are believed to be powered by mass accretion onto black holes. Their emitting range is from $10^{40}$ erg s$^{-1}$ (e.g., the nuclei of some nearby galaxies) to more than $10^{47}$ erg s$^{-1}$ (distant quasars), which is far beyond the Eddington luminosity (about $10^{44}$ erg s$^{-1}$).
Therefore, the X-ray variability in AGNs and the disc emission lines might contain imprints of the radiation drag effects.

In a series of papers we have studied the dynamics of test particles in a given gravitational background while subject to a Thomson-type interaction with a superimposed test radiation field, namely the so called Poynting-Robertson (PR) effect \citep{poynting,robertson}. 
We modeled the radiation flux as made of photons emitted all along a common direction from a point-like source, leading to an effective drag force to be added to the gravitational force. 
More realistic descriptions would take into account the finite size of the radiating source and allow for photons to be emitted in any direction. However, the computation of the radiation force in this case usually requires complicated numerical ray-tracing calculations for null geodesics in the background spacetime. The general relativistic stress-energy tensor associated with such a more general scenario was constructed for the first time in the pioneering work of \citet{abram}, who considered the special case of test bodies radially moving in the spacetime of a spherically symmetric non-rotating radiation source with finite radius to model jets and solar winds. They studied the radial equilibrium solutions in which the test particle remains at rest under the combined inward gravitational attraction and the outward photon pressure (the \lq\lq Eddington sphere''). The stability of this equilibrium configuration has been recently investigated by \citet{stahl}, who have also discussed its implications for Hoyle-Lyttleton accretion onto a luminous star (see also \citep{oh}).   
\citet{wielgus}, \citet{stahl2} and \citet{mishra} have then analyzed the effects of a luminosity variation on the Eddington sphere, especially in view of the possible occurrence of coronal ejection from the system.
Within this framework, leaving the spherical symmetry makes the problem very difficult to be addressed. 
The generalization of the Abramowicz et al. approach to the case of an arbitrary particle motion in the equatorial plane of a rotating source is due to \citet{miller}, but in the limit of slow rotation only. 
Their model was then used by \citet{oh2} to study the existence of equilibrium solutions.

Following the original Robertson approach, we have considered so far different spacetime solutions of astrophysical interest endowed with reflection symmetry with respect to the equatorial plane, like black hole solutions (Schwarzschild and Kerr) \citep{bijanste,PR2,PRfala} or radiating spacetimes (Vaidya) \citep{PRvaidya}.
We have also discussed the effect of coupling with additional properties of test particles (like intrinsic spin) \citep{PRspin}.
The main result of our previous analysis is that particles moving on the symmetry plane (which do not escape) are definitely attracted to a certain critical radius where they orbit the central source maintaining equilibrium. 
In general, the critical radius is not unique for sufficiently large values of the impact parameter of the photons as well as of the strength of the radiation field.

Here we explore the case of a central source having a quadrupolar structure as described by the Erez-Rosen solution. 
This is an exact solutions to the vacuum Einstein's field equations which generalizes the Schwarzschild spacetime to the case of a gravitational source endowed with an arbitrary mass quadrupole moment, and hence it is specified by two parameters, the mass $M$ and the quadrupole parameter $q$.
This situation has a strong astrophysical motivation and represents a natural generalization of previous results, even if rotational effects are neglected.
In fact, also the Kerr solution possesses a non-vanishing quadrupole moment, but the latter is induced by the spacetime rotation.
Here, the quadrupole parameter $q$ describes a genuine mass quadrupole moment directly related to the shape deformation of the source. 
For instance, the supermassive objects hosted in galaxy centers are generally expected to be also endowed with a non-negligible mass quadrupole moment, which may affect the dynamics of stars moving in the region very close to the central object \citep{will}.
Besides rotation, torques from the source's quadrupole moment cause in general precession of the stellar orbital planes, which can be measured by high-precision astrometry. This is the case of the compact cluster of stars orbiting the center of the Milky Way galaxy at milliparsec distances, monitored since many years \citep{gillessen}.

Taking into account quadrupolar deformations is also important when considering the spacetime region around neutron stars, which can assume a pronounced oblate configuration. Neutron stars can be part of low mass X-ray binaries (LMXBs), where they are spun up by matter that accretes from a companion, which is typically either a low mass star or a white dwarf. For a dynamically negligible magnetic field, the accretion disc of LMXBs can extend down to the star's surface, where curvature effects dominate.
Therefore, the observed flux emitted in the inner part of the disc may contain the signatures of a more complicated structure of the source, which can be inferred from available X-ray data.
However, significant quadrupolar deformations are mainly due to rotation, as in the case of Kerr black holes or rapidly rotating neutron stars.
Hence, an accurate description of the above systems would need also the inclusion of rotation, which is but beyond the scope of the present analysis, mostly motivated to better understand the role of the quadrupole. 

The main outcome of this work is the analysis of the effects induced on the equilibrium solutions by the quadrupolar deformation of the source.
This is a typical feature of particles undergoing PR effect already discussed in previous works, as already mentioned.
We will show that the presence of the quadrupolar structure of the source of the gravitational field may affect the conditions for equilibrium in a significant way, introducing an additional degree of freedom with respect to spherically symmetric sources represented by the quadrupole parameter.
This feature can in principle give rise to observable effects. 

We use geometrical units ($c=G=1$) and follow notations and conventions of \citet{MTW}.
The metric signature is $+2$.
Latin indices run from 1 to 3, greek indices from 0 to 3.

\section{Test particles undergoing PR effect in the Erez-Rosen spacetime} 

The gravitational field of a nonrotating mass with a quadrupole moment can be described by the Erez-Rosen solution \citep{erez,novikov,young}.
It belongs to the static Weyl class of solutions with the line element written in prolate spheroidal coordinates ($t,x,y,\phi$), with $x \geq 1$ and $-1 \leq y \leq 1$, as follows \citep{ES}
\beq
\label{metric_Weyl}
\rmd s^2=-f\rmd t^2+\frac{\sigma^2}{f}\left\{e^{2\gamma}\left(x^2-y^2\right)\left(\frac{\rmd x^2}{x^2-1}+\frac{\rmd y^2}{1-y^2}\right)+(x^2-1)(1-y^2)\rmd\phi^2\right\}\,,
\eeq
where $\sigma$ is a constant and the quantities $f$ and $\gamma$ are functions of $x$ and $y$ only.
The metric functions are given by
\begin{eqnarray}
\label{metdef}
f&=&\frac{x-1}{x+1}e^{-2qP_2Q_2}\,, \nonumber\\
\gamma&=&\frac12(1+q)^2 \ln\frac{x^2-1}{x^2-y^2} + 2q(1-P_2)Q_1 + q^2(1-P_2) \bigg[ (1+P_2)(Q_1^2-Q_2^2)\nonumber \\
&&+\frac12(x^2-1)(2Q_2^2 - 3xQ_1Q_2 + 3 Q_0Q_2 - Q_2')\bigg] \,,
\end{eqnarray}
where $P_l(y)$ and $Q_l(x)$ are Legendre polynomials of the first and second kind, respectively (see Appendix A), and $q$ is the dimensionless quadrupole parameter. 
Positive values of $q$ correspond to prolate configurations, i.e., the mass is mostly concentrated along the axes $y=\pm1$, whereas negative values to oblate ones.
When $q=0$, the metric (\ref{metric_Weyl}) reduces to the Schwarzschild solution provided that $\sigma$ be identified with the mass of the source, namely $\sigma=M$.
Transition of this  metric form to the more familiar one associated with standard Schwarzschild-like coordinates is accomplished by the following coordinate transformation $x={r}/{M}-1$ and $y=\cos\theta$.
Furthermore, the above solution reduces to the well known Hartle-Thorne spacetime \citep{HT} with vanishing rotation parameter when linearized with respect to the quadrupole parameter.

A suitable family of fiducial observers is that of the so called static observers, with unit timelike four velocity $n\equiv e_{\hat t}= (1/\sqrt{f})\partial_t\,$ aligned with the timelike Killing vector $\partial_t$.
They are accelerated with acceleration $a(n)=\nabla_n n$.
An orthonormal frame adapted to the static observers is thus given by
\beq
e_{\hat t}=n , \,\quad
e_{\hat x}=\frac1{\sqrt{g_{xx}}}\partial_x\,,\qquad
e_{\hat y}=\frac1{\sqrt{g_{yy}}}\partial_y\,,\quad
e_{\hat \phi}=\frac1{\sqrt{g_{\phi \phi }}}\partial_\phi\,.
\eeq

The four acceleration as well as the curvature vectors $k(x^i,n)$, where $x^i=x,y,\phi$, associated with the diagonal metric coefficients
\citep{mfg,idcf12,bjdf} only have nonzero components in the $x$-$y$ 2-plane of the tangent space, i.e.,
\begin{eqnarray}
\label{accexp}
a(n) & = & a(n)^{\hat x} e_{\hat x} + a(n)^{\hat y} e_{\hat y}
 =\partial_{\hat x}(\ln\sqrt{f}) e_{\hat x} + \partial_{\hat y}(\ln\sqrt{f})  e_{\hat y}
\,,
\nonumber\\
k(x^i,n)
& = &k(x^i,n)^{\hat x} e_{\hat x} + k(x^i,n)^{\hat y} e_{\hat y}
 = -\partial_{\hat x}(\ln \sqrt{g_{ii}}) e_{\hat x} - \partial_{\hat y}(\ln \sqrt{g_{ii}})e_{\hat y}
\,.
\end{eqnarray}
We limit our analysis to the symmetry plane $y=0$.
Therefore, both the acceleration and the curvature vectors are directed along the $x-$axis, which hereafter will be referred to as the \lq\lq radial'' direction.

\subsection{Photon field}

Let a pure electromagnetic radiation field be superposed as a test field on the gravitational background described by the metric (\ref{metric_Weyl})--(\ref{metdef}),  with the energy-momentum tensor
\beq
\label{ten_imp}
T^{\alpha\beta}=\Phi^2 k^\alpha k^\beta, \qquad k^\alpha k_\alpha=0\ ,
\eeq
where $k$ is assumed to be tangent to an affinely parametrized (ingoing/outgoing) null geodesic in the symmetry plane, i.e., $k^\alpha \nabla_\alpha k^\beta=0$ with $k^y=0$.
We then have
\beq
\label{kdef}
k=E(n)[n+\hat \nu(k,n)]\,, \qquad
\hat \nu(k,n)=\sin \beta\, e_{\hat x}+\cos \beta\, e_{\hat \phi}\,,
\eeq
where 
\beq
E(n)= - k \cdot n
=\frac{E}{\sqrt{f}}
\eeq
is the relative energy of the photons and
\beq
\label{cosbeta}
\cos \beta 
=\frac{bf}{\sigma\sqrt{x^2-1}}\,.
\eeq 
The constant $b=L/E$ denotes the photon impact parameter defined in terms of the conserved energy $E=-k_t>0$ and angular momentum $L=k_\phi$ associated with the timelike and azimuthal Killing vectors, respectively.
The case $\sin \beta >0$ corresponds to outgoing photons (increasing radial distance from the central source) and $\sin \beta <0$ to incoming photons (decreasing $x$). 

Since $k$ is completely determined,  the coordinate dependence of the quantity $\Phi$  then follows 
from the conservation equations  $T^{\alpha\beta}{}_{;\beta}=0$, and will only depend on $x$ in the symmetry plane.
From Eq.~(\ref{ten_imp}) using the geodesic condition for $k$, these can be written as
\beq
\label{flux_cons}
0=\nabla_\beta (\Phi^2 k^\beta)
=\frac{1}{\sqrt{-g}}\partial_\beta (\sqrt{-g}\,\Phi^2 k^\beta)
\,, 
\eeq
implying that $0=\partial_x(\sqrt{-g}\,\Phi^2 k^x)$.
Therefore, we find $\sqrt{-g}\,\Phi^2 k^x=\hbox{\rm const} = E \Phi_0^2 $, leading to  
\beq
\Phi^2=\sigma\Phi_0^2\frac{e^{-\gamma}}{x\,|b \tan \beta|} 
=\sigma\Phi_0^2\frac{e^{-\gamma}f}{x\sqrt{\sigma^2(x^2-1)-b^2f^2}} \,.
\eeq
In the limit $b\to0$ corresponding to photons in radial motion the flux simplifies as
\beq
\Phi^2=\Phi_0^2\frac{e^{-\gamma}f}{x\sqrt{x^2-1}} \,.
\eeq

\begin{figure}
\centering
\subfigure{\includegraphics[scale=0.23]{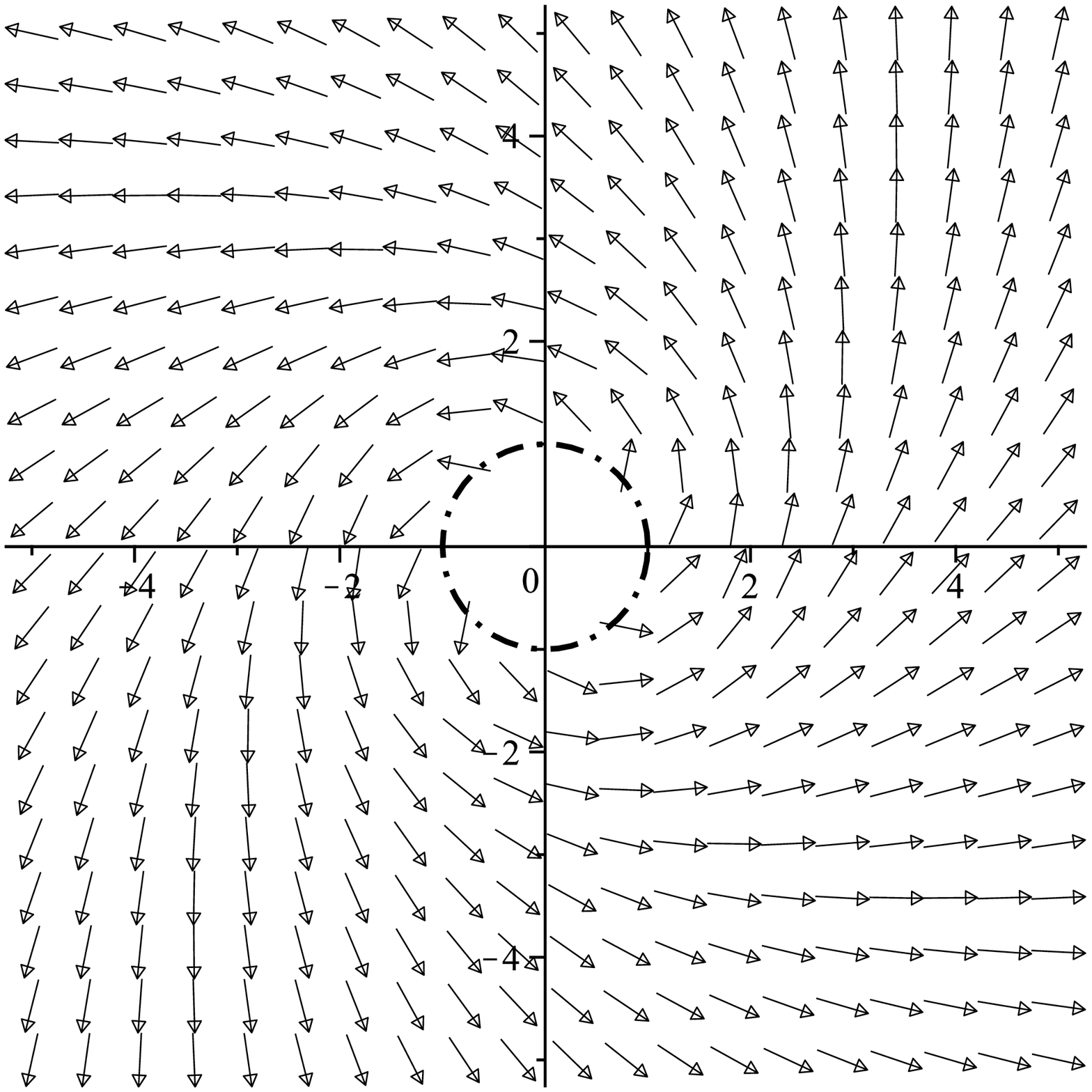}}
\hspace{5mm}
\subfigure{\includegraphics[scale=0.25]{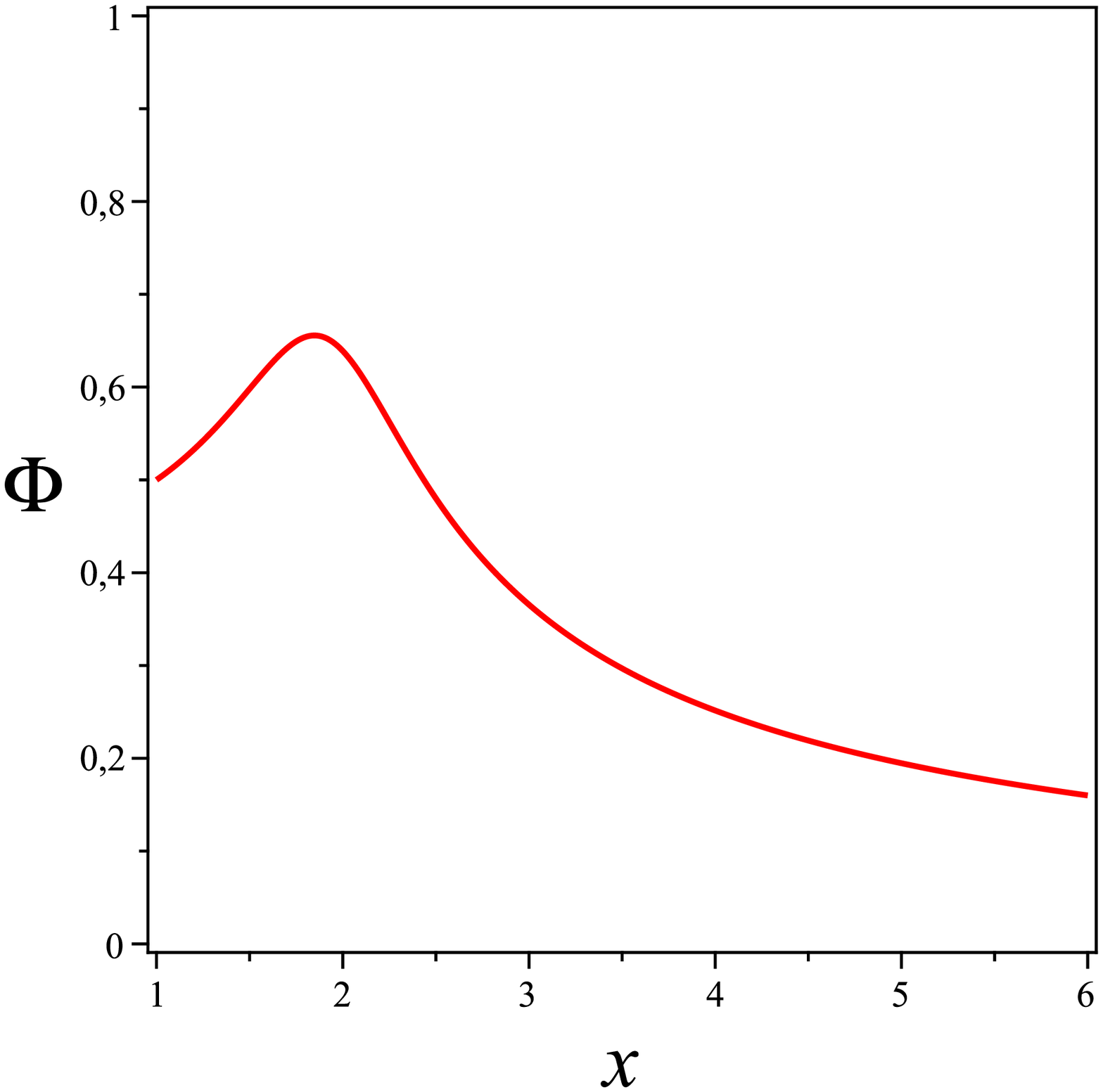}}\\[1cm]
\subfigure{\includegraphics[scale=0.23]{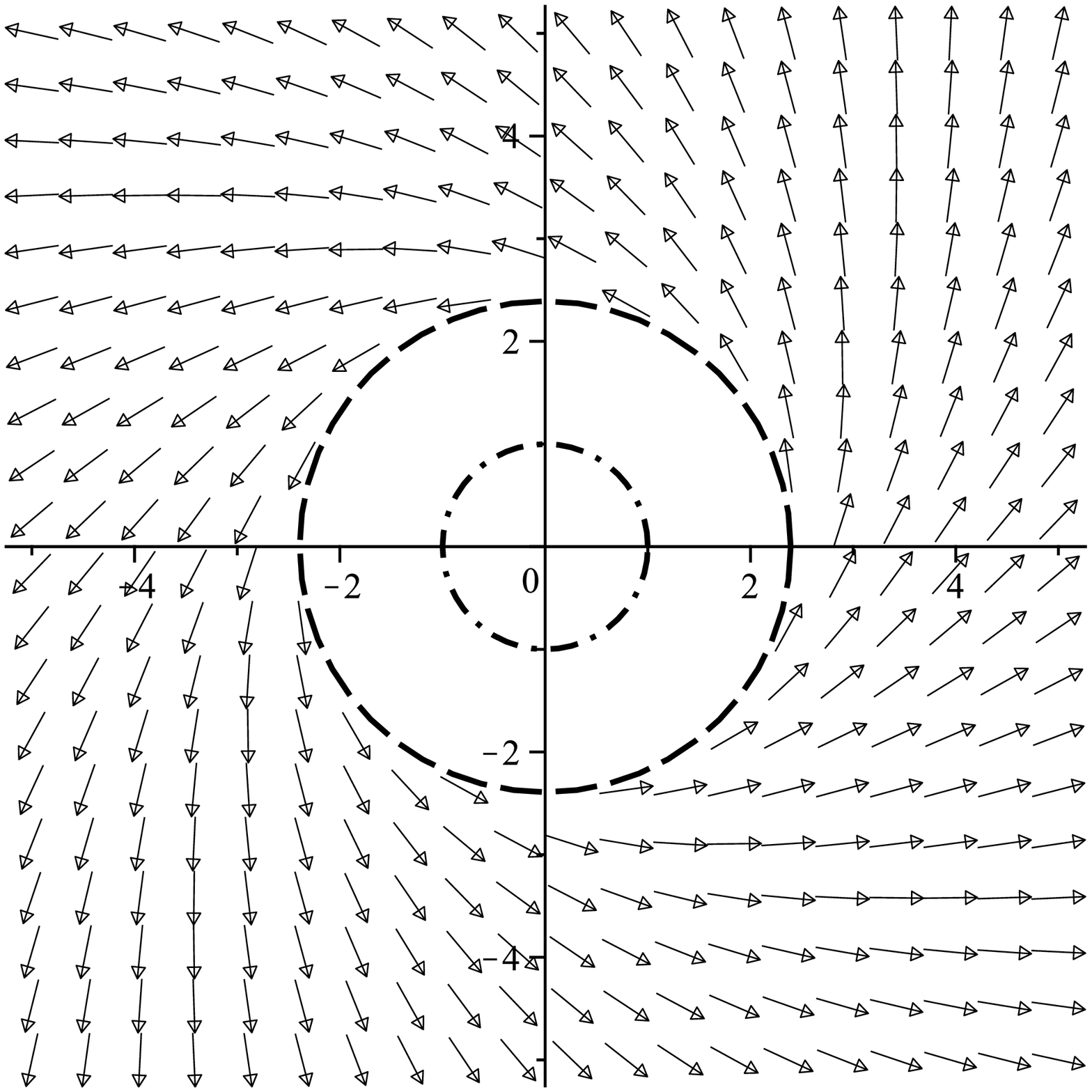}}
\hspace{5mm}
\subfigure{\includegraphics[scale=0.25]{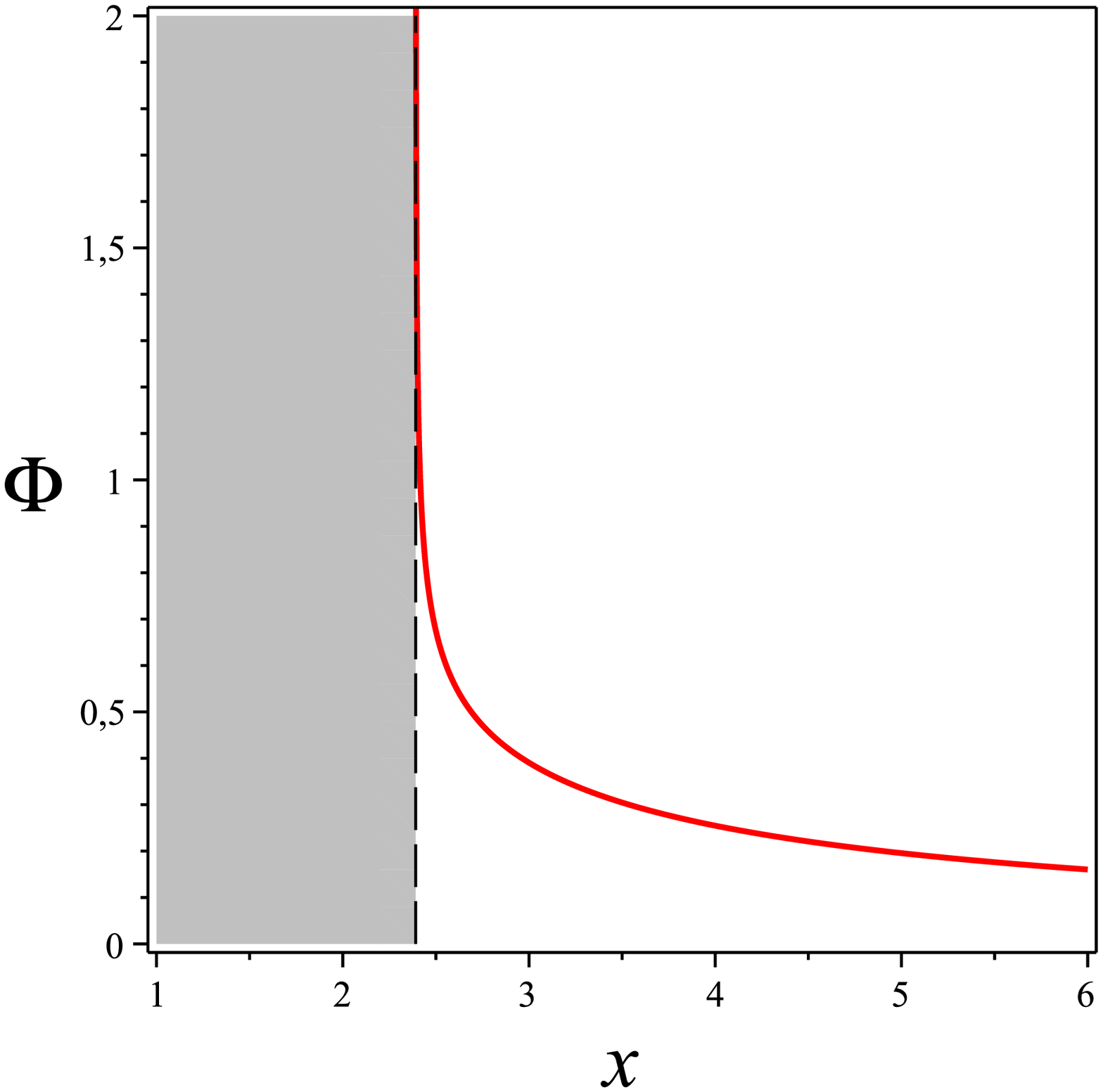}}
\caption{
{\bf First row}: The unit velocity direction field of photons $\hat \nu(k,n)$ given by Eq. (\ref{kdef}) is shown in the left panel for $b/\sigma=5$ and $q=0$.
The right panel shows instead the behavior of the photon flux (in units of $\Phi_0$) for the same choice of parameters.
{\bf Second row}: 
Same value of the photon impact parameter as above, but with nonzero quadrupole parameter $q=5$.
The region inside the dashed circle at $x\approx2.39$ is forbidden to photons due to the restriction $\cos\beta\le1$ as from Eq. (\ref{cosbeta}). 
}
\label{fig:flux}
\end{figure}

The photon motion in the symmetry plane has been investigated by \citet{ERlight} through the analysis of the associated effective potential.
A number of interesting features has been discussed there, including the occurrence of multiple \lq\lq photon spheres'' corresponding to spatially circular null orbits due to the presence of a nonvanishing mass quadrupole moment.
Fig. \ref{fig:flux} shows the behavior of the relative velocity lines of the radiation field (with respect to the observers $n$) as well as that of the radiation flux as a function of the radial distance in the case of a Erez-Rosen spacetime with a selected value of the quadrupole parameter in comparison with the spherically symmetric Schwarzschild solution.
In the latter case the flux is peaked at $x=2$ (i.e., $r=3M$) and approaches a finite value at the horizon, for the chosen (high) value of the photon impact parameter.
In the case of a non-spherical source, instead, the behavior of the flux strongly depends on $q$. 
Fig. \ref{fig:flux} shows a typical situation corresponding to a prolate configuration ($q=5$).
As it is evident, there exists a forbidden region to photons extending from $x=1$ up to a certain value of $x$, where the flux diverges, hence manifesting a completely different behavior in comparison with the Schwarzschild case.
For oblate configurations the situation is instead similar to the spherically symmetric case.


\subsection{Particle dynamics}

Consider now a test particle moving in the symmetry plane $y=0$ accelerated by the radiation field, i.e. with 4-velocity 
\beq\label{polarnu}
U=\gamma(U,n) [n+ \nu(U,n)]\,,\quad 
\nu(U,n)\equiv \nu^{\hat x}e_{\hat x}+\nu^{\hat \phi}e_{\hat \phi} 
\,,
\eeq
where $\gamma(U,n)=1/\sqrt{1-||\nu(U,n)||^2}\equiv\gamma$ is the Lorentz factor (not to be confused with the metric function) and the abbreviated notation $\nu^{\hat a}\equiv\nu(U,n)^{\hat a}$ has been used.  
A straightforward calculation gives the coordinate components of $U$
\beq
\label{Ucoord_comp}
U^t\equiv \frac{\rmd t}{\rmd \tau}=\frac{\gamma}{\sqrt{f}}\,, \qquad 
U^x\equiv \frac{\rmd x}{\rmd \tau}=\frac{\gamma\nu^{\hat x}}{\sqrt{g_{xx}}}\,, \qquad 
U^\phi\equiv \frac{\rmd \phi}{\rmd \tau}=\frac{\gamma\nu^{\hat \phi}}{\sqrt{g_{\phi\phi}}}\,,
\eeq
where $\tau$ is the proper time parameter along the particle's world line, and $U^y\equiv \rmd y/\rmd \tau=0$. 

The scattering of radiation as well as the (constant) momentum-transfer cross section $\tilde \sigma$ of the particle are assumed to be independent of the direction and frequency of the radiation so that the associated force is given by  \citep{poynting,robertson}
\beq
\label{frad}
{\mathcal F}_{\rm (rad)}(U)^\alpha = -\tilde \sigma P(U)^\alpha{}_\beta \, T^{\beta}{}_\mu \, U^\mu\,,
\eeq
where $P(U)^\alpha{}_\beta=\delta^\alpha_\beta+U^\alpha U_\beta$ projects orthogonally to $U$.
Explicitly 
\beq
{\mathcal F}_{\rm (rad)}(U)^\alpha=-\tilde \sigma \Phi^2 [P(U)^\alpha{}_\beta k^\beta]\, (k_\mu U^\mu)
=\tilde \sigma \, [\Phi E(U)]^2\, \hat {\mathcal V}(k,U)^\alpha\,,
\eeq
where 
\beq
E(U)=-U\cdot k
= \gamma E(n)[1-\sin\beta\nu^{\hat x}-\cos\beta\nu^{\hat \phi}]\,,
\eeq
is the photon energy as measured by $U$ and 
\beq
\hat {\mathcal V}\equiv\hat {\mathcal V}(k,U)
=\frac{k}{E(U)}-U\,,
\eeq
is the photon relative velocity in the test particle local rest space with the property $\hat{\mathcal V}\cdot U=0$.
The decomposition of the radiation force with respect to $n$ is thus given by
\beq
{\mathcal F}_{\rm (rad)}(U)={\mathcal F}_{\rm (rad)}(U)^{\hat t}n
+{\mathcal F}_{\rm (rad)}(U)^{\hat x}e_{\hat x}
+{\mathcal F}_{\rm (rad)}(U)^{\hat \phi}e_{\hat \phi}\,,
\eeq
with components ${\mathcal F}_{\rm (rad)}(U)^{\hat t}={\mathcal F}_{\rm (rad)}(U)^{\hat x} \nu^{\hat x} + {\mathcal F}_{\rm (rad)}(U)^{\hat \phi} \nu^{\hat \phi}$ and 
\begin{eqnarray}
{\mathcal F}_{\rm (rad)}(U)^{\hat x}
& = &\tilde \sigma \, [\Phi E(U)]^2\, \hat {\mathcal V}^{\hat x}\,, 
\nonumber\\
{\mathcal F}_{\rm (rad)}(U)^{\hat \phi}
& = &\tilde \sigma \, [\Phi E(U)]^2\, \hat {\mathcal V}^{\hat \phi} 
\,,
\end{eqnarray}
with 
\begin{eqnarray}
\label{eq:hatVu}
\hat {\mathcal V}{}^{\hat x}&=&\sin \beta\frac{E(n)}{E(U)} -\gamma\nu^{\hat x}\,,\nonumber \\
\hat {\mathcal V}{}^{\hat \phi}&=&\cos \beta\frac{E(n)}{E(U)} -\gamma\nu^{\hat \phi}\,.
\end{eqnarray}

Test particle motion is then described by the equation
\beq
\label{eqPRtest}
m a(U)^\mu\equiv m \frac{\rm{D} U^\mu}{\rmd\tau}={\mathcal F}_{\rm (rad)}(U)^\mu\,.
\eeq
The frame components of the particle 4-acceleration $a(U)$ in the symmetry plane are given by
\begin{eqnarray}
\label{eq_fundam}
a(U)^{\hat t}
&=&a(U)^{\hat x} \nu^{\hat x} +a(U)^{\hat \phi} \nu^{\hat \phi}\,,\nonumber \\
\gamma^{-2}a(U)^{\hat x}
&=&  a(n)^{\hat x}+k(\phi,n)^{\hat x}\,\nu^{\hat \phi}{}^2
+\gamma\left[\nu^{\hat x}\nu^{\hat \phi}\frac{\rmd \nu^{\hat \phi}}{\rmd \tau} 
       +(1-\nu^{\hat \phi}{}^2)\frac{\rmd \nu^{\hat x}}{\rmd \tau}\right]\,, \nonumber \\
a(U)^{\hat y}
&=& 0\,, \nonumber \\
\gamma^{-2}a(U)^{\hat \phi}
&=& -\nu^{\hat x}\nu^{\hat \phi}\, k(\phi,n)^{\hat x} 
+\gamma\left[\nu^{\hat x}\nu^{\hat \phi}\frac{\rmd \nu^{\hat x}}{\rmd \tau} 
       +(1-\nu^{\hat x}{}^2)\frac{\rmd \nu^{\hat \phi}}{\rmd \tau}\right]\,.
\end{eqnarray}
From Eq. (\ref{eqPRtest}), the evolution equations for the frame components of the linear velocity then read as
\begin{eqnarray}
\label{motioneqs}
\frac{\rmd \nu^{\hat x}}{\rmd \tau}
&=&-\gamma(1-\nu^{\hat x}{}^2)\left[a(n)^{\hat x}-\frac{{\mathcal F}_{\rm (rad)}(U)^{\hat x}}{m\gamma^2}\right]
-\gamma\nu^{\hat \phi}\left[\nu^{\hat \phi}k(\phi,n)^{\hat x}+\nu^{\hat x}\frac{{\mathcal F}_{\rm (rad)}(U)^{\hat \phi}}{m\gamma^2}\right]\,, \nonumber \\
\frac{\rmd \nu^{\hat \phi}}{\rmd \tau}
&=&(1-\nu^{\hat \phi}{}^2)\frac{{\mathcal F}_{\rm (rad)}(U)^{\hat \phi}}{m\gamma}
+\gamma\nu^{\hat x}\nu^{\hat \phi}\left[a(n)^{\hat x}+k(\phi,n)^{\hat x} -\frac{{\mathcal F}_{\rm (rad)}(U)^{\hat x}}{m\gamma^2}\right]\,,
\end{eqnarray}
which should be considered together with Eq. (\ref{Ucoord_comp}).
The explicit expressions for the radial components of the acceleration as well as curvature vectors are listed in Appendix A.

\begin{figure}
\centering
\subfigure{\includegraphics[scale=0.2]{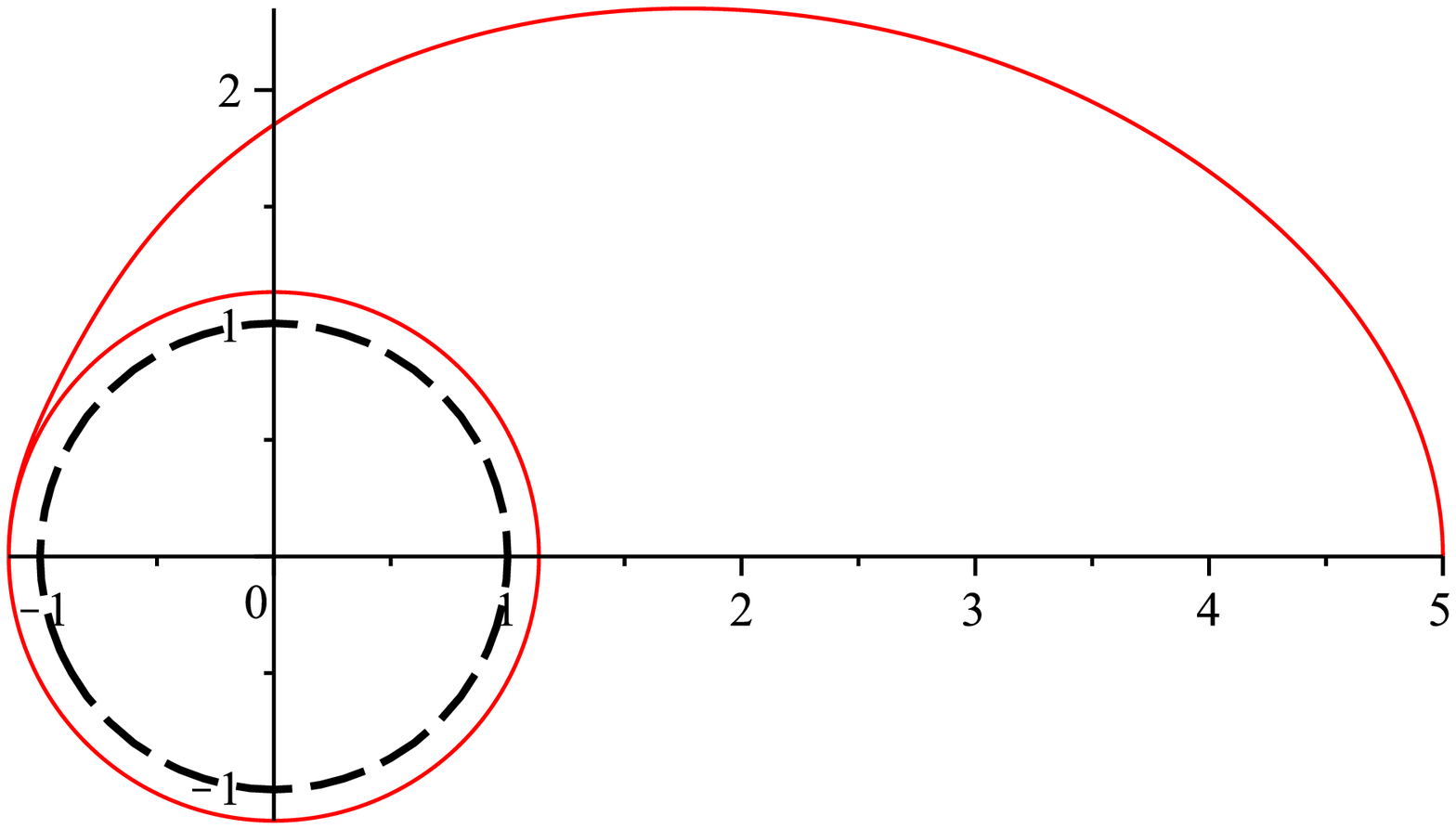}}
\hspace{5mm}
\subfigure{\includegraphics[scale=0.2]{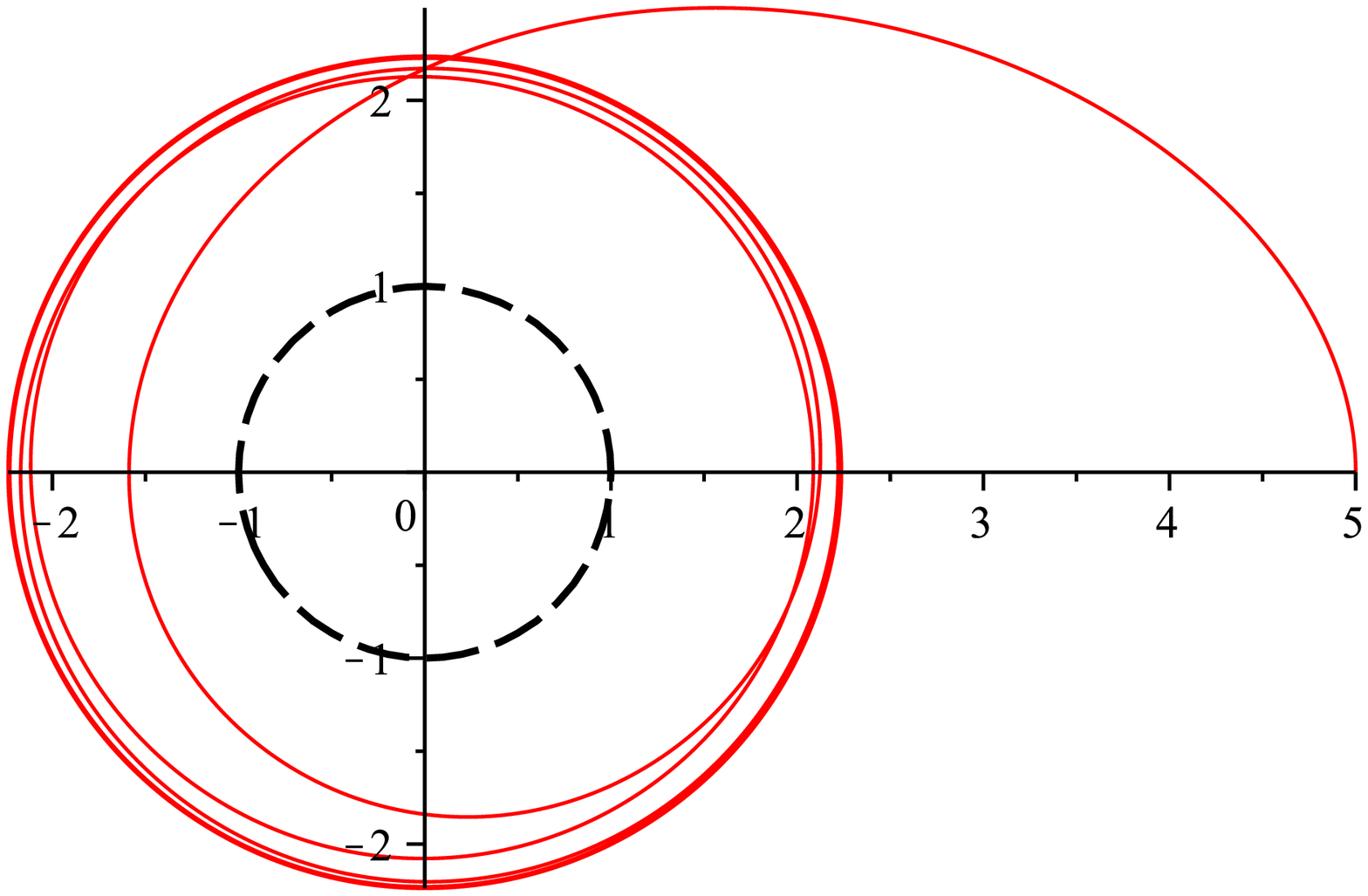}}
\hspace{5mm}
\subfigure{\includegraphics[scale=0.2]{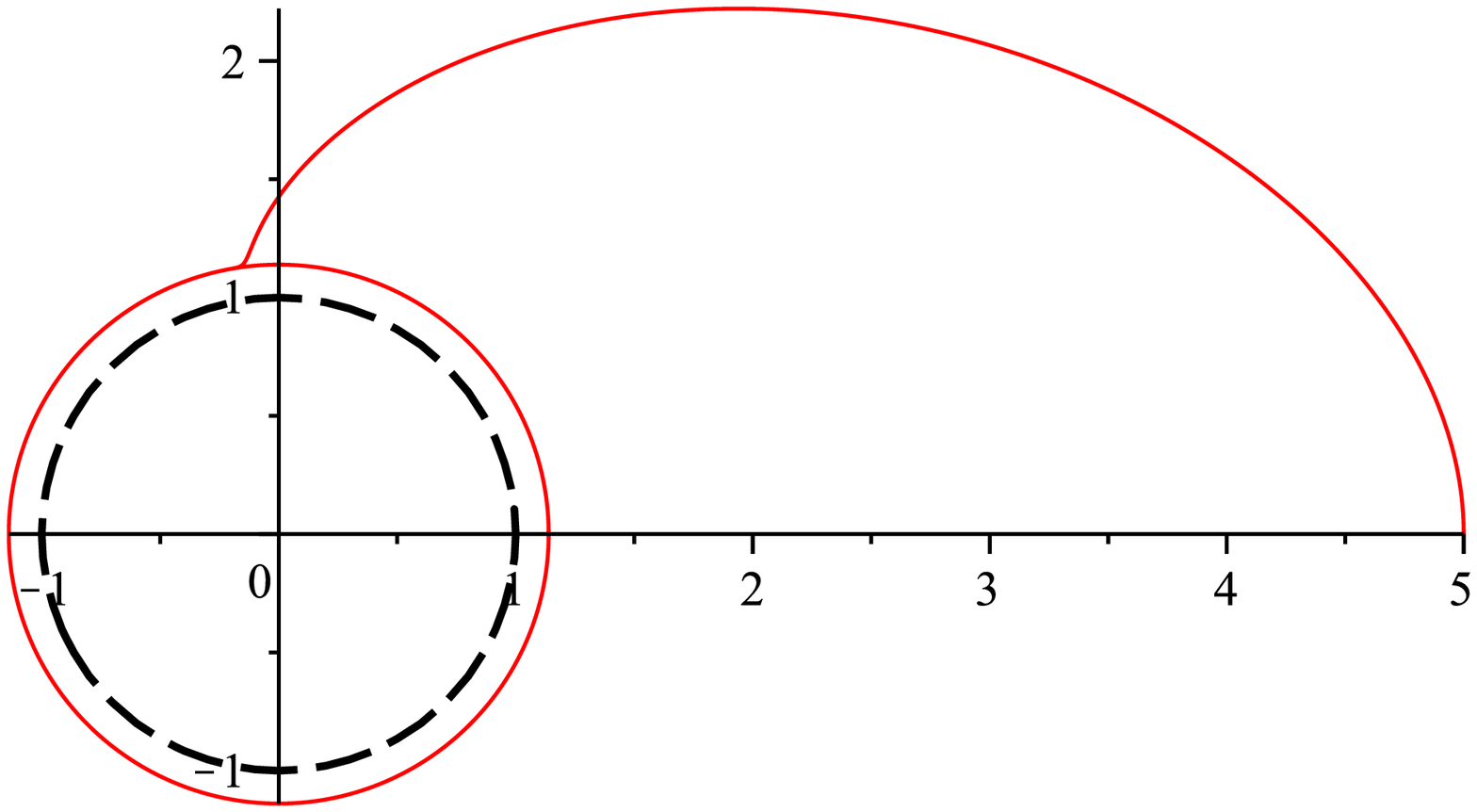}}\\[1cm]
\subfigure{\includegraphics[scale=0.2]{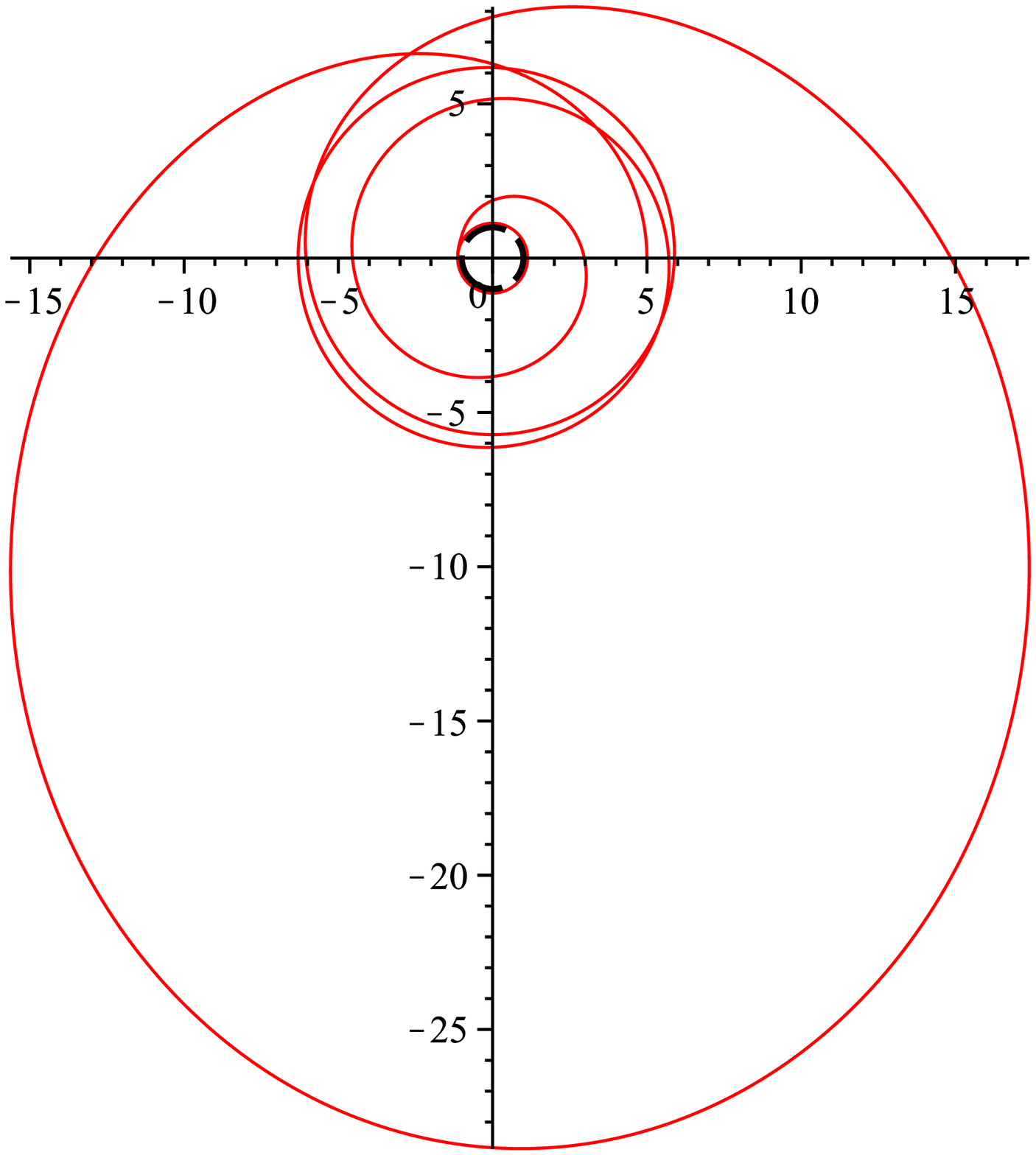}}
\hspace{5mm}
\subfigure{\includegraphics[scale=0.2]{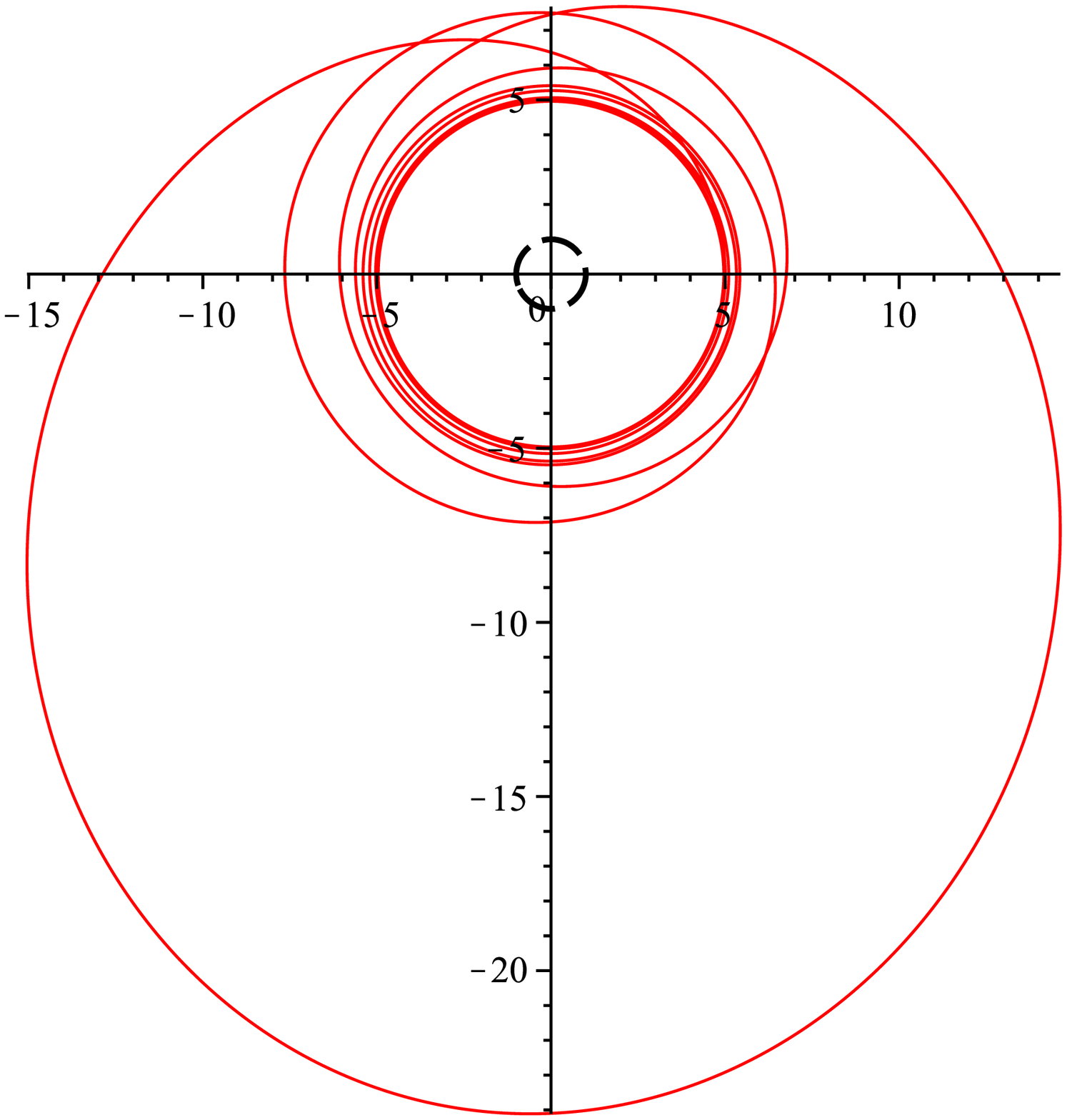}}
\hspace{5mm}
\subfigure{\includegraphics[scale=0.2]{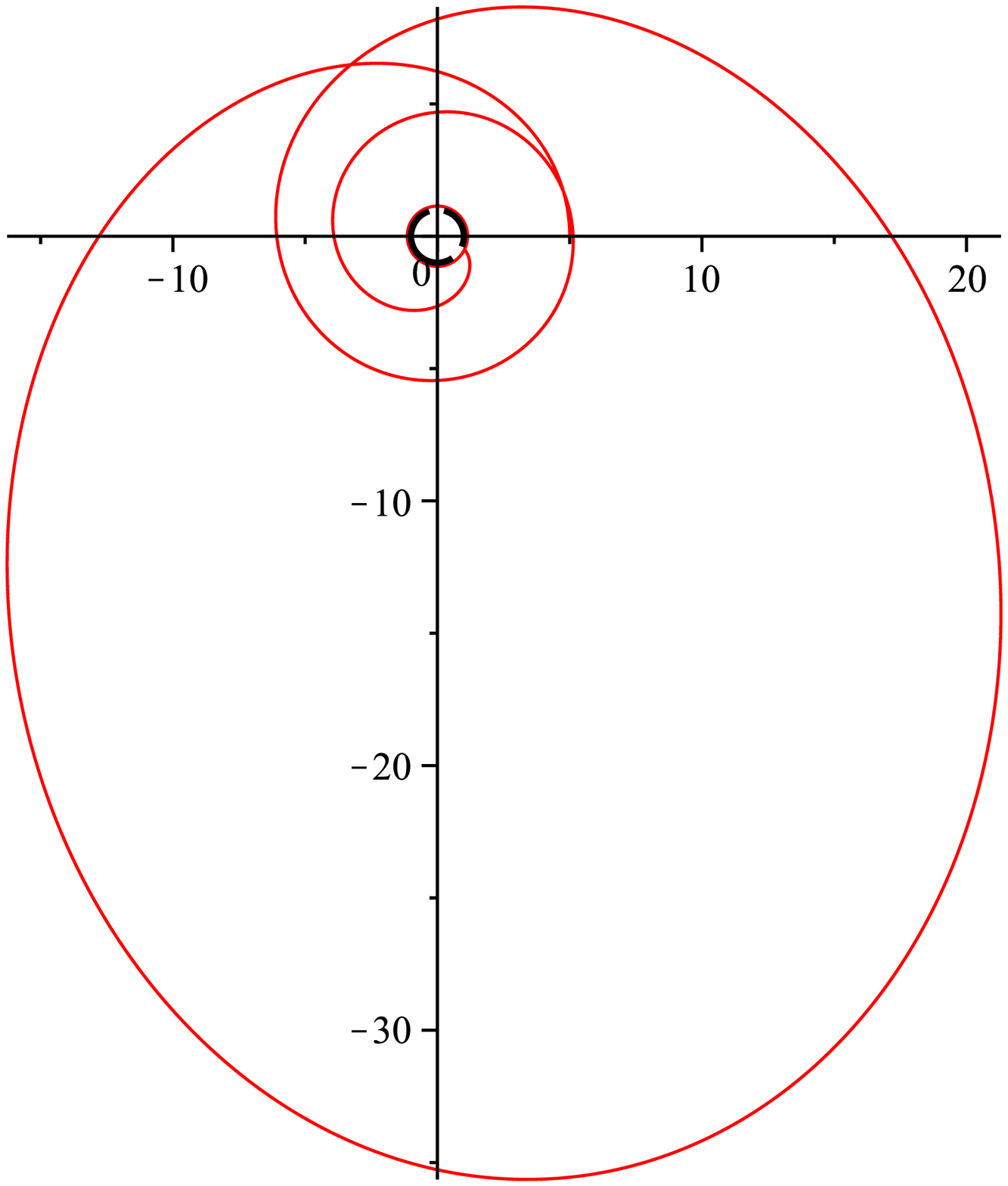}}
\caption{
{\bf First row}: Particle orbits on the symmetry plane for $b=3$, $A=0.3$ (in units of $\sigma$) and different values of the quadrupole parameter: $q=0$ (left), $q=5$ (middle) and $q=-5$ (right).
Initial conditions have been chosen as $x(0)=5$, $\phi(0)=0$, $\nu^{\hat r}(0)=0$ and $\nu^{\hat\phi}(0)=0.2$. 
Cartesian-like coordinates $[x\cos\phi,x\sin\phi]$ have been used.
The equilibrium circular orbits are at $x\approx1.134$, $x\approx2.238$ and $x\approx1.139$, respectively.
{\bf Second row}: Same parameter choice as above, but with different $\nu^{\hat\phi}(0)=\nu_{\rm g}\approx[0.5,0.487,0.513]$, respectively.
The equilibrium circular orbits are at $x\approx1.134$, $x\approx4.962$ and $x\approx1.139$, respectively.
}
\label{fig:orbite}
\end{figure}

Examples of numerical integration of the orbits are shown in Fig. \ref{fig:orbite} for a fixed value of the radiation field strength and different values of the quadrupole parameter.
The case of a spherically symmetric source ($q=0$) is also shown for comparison.
Initial conditions are chosen so that the particle trajectory is initially tangent to a circular orbit with a value of the azimuthal velocity which is either less than (first row) or equal to (second row) the geodesic one.
As a general feature, unless the particle has a sufficiently high initial speed that it can escape to infinity, it is forced to end up in a circular orbit which is an equilibrium solution, representing the balance between gravitational attraction, centrifugal force and radiation drag, as discussed in the next section.
Fig. \ref{fig:orbite} shows that particles feel a gravitational field with different strength if the source is oblate ($q<0$) or prolate ($q>0$).
In the former case the incoming particle feels a stronger gravitational field as it approaches the gravity source.
For prolate configurations, instead, the gravitational mass is mainly concentrated along the $y-$axis, so that a particle moving sufficiently close to the source will feel a decreasing gravitational field.
As a result, the balance between inward gravitational force and outward radiation force will occur at larger radii.


\section{Equilibrium solutions}

The set of equations (\ref{motioneqs}) and (\ref{Ucoord_comp}) admits an equilibrium solution which corresponds to a circular orbit with constant speed at a given distance $x=x_0$ from the central source.  
In fact, setting $\nu^{\hat x}=0$ and $\nu^{\hat\phi}=\nu^{\hat\phi}_0=$ const. in the above equations we get the conditions
\begin{eqnarray}
\label{equileqs}
0&=& a(n)^{\hat x}+\nu^{\hat \phi}_0{}^2k(\phi,n)^{\hat x}-\frac{A}{\sigma^2}\frac{\Phi^2}{\Phi_0^2}\frac{\sin\beta_0}{f\gamma_0}(1-\cos\beta_0\nu^{\hat \phi}_0)
\,,\nonumber\\
0&=&A\frac{\Phi^2}{\Phi_0^2}\frac{\gamma_0^3}{f}(\cos\beta_0 -\nu^{\hat\phi}_0)(1 -\nu^{\hat\phi}_0\cos \beta_0 )
\,,
\end{eqnarray}
where it is understood that all functions of $x$ are evaluated at $x=x_0$ and we have used the notation $\tilde \sigma \Phi_0^2 E^2=mA$, as in \citep{PR2}.
The second Eq.~(\ref{equileqs}) yields the critical circular velocity  
\beq
\label{nu0limit}
\nu^{\hat\phi}_0 = \cos \beta_0 
       = \frac{bf}{\sigma\sqrt{x^2-1}} \qquad 
\rightarrow \qquad
  \gamma_0= \frac{1}{|\sin\beta_0|}
\,,
\eeq
which turns out to be equal to the photon azimuthal velocity (see Eq. (\ref{cosbeta})).
Substituting then into the first equation gives an implicit relation determining the critical \lq\lq radius'' $x_0$ itself as a function of $b$, which can be also written as 
\beq
\label{equil}
a(n)^{\hat x}\left(1-\frac{\nu^{\hat \phi}_0{}^2}{\nu_{\rm g}^2}\right)=\frac{A}{\sigma^2}\frac{\Phi^2}{\Phi_0^2}\frac{{\rm sgn}(\sin \beta_0)}{f\gamma_0^4}
\,,
\eeq
where the circular geodesic ($A=0$) azimuthal velocities $\nu^{\hat \phi}=\pm\nu_{\rm g}$, with
\beq
\nu_{\rm g}=\left[\frac{f_x(x^2-1)}{2xf-f_x(x^2-1)}\right]^{1/2}\,, \qquad
\gamma_{\rm g}=\frac1{\sqrt{1-\nu_{\rm g}^2}}\,,
\eeq
have been introduced, satisfying the relation $a(n)^{\hat x}+\nu_{\rm g}^2k(\phi,n)^{\hat x}=0$.
Partial derivatives of the metric functions with respect to $x$ and $y$ are indicated by a subscript.
Solving Eq. (\ref{equil}) for $A$ finally gives
\beq
\label{equil2}
\frac{A}{\sigma}={\rm sgn}(\sin \beta_0)\gamma_0^3\left(1-\frac{\nu^{\hat \phi}_0{}^2}{\nu_{\rm g}^2}\right)\frac{f_x}{2\sqrt{f}}(x^2-1)
\,.
\eeq

\begin{figure}
\centering
\subfigure{\includegraphics[scale=0.2]{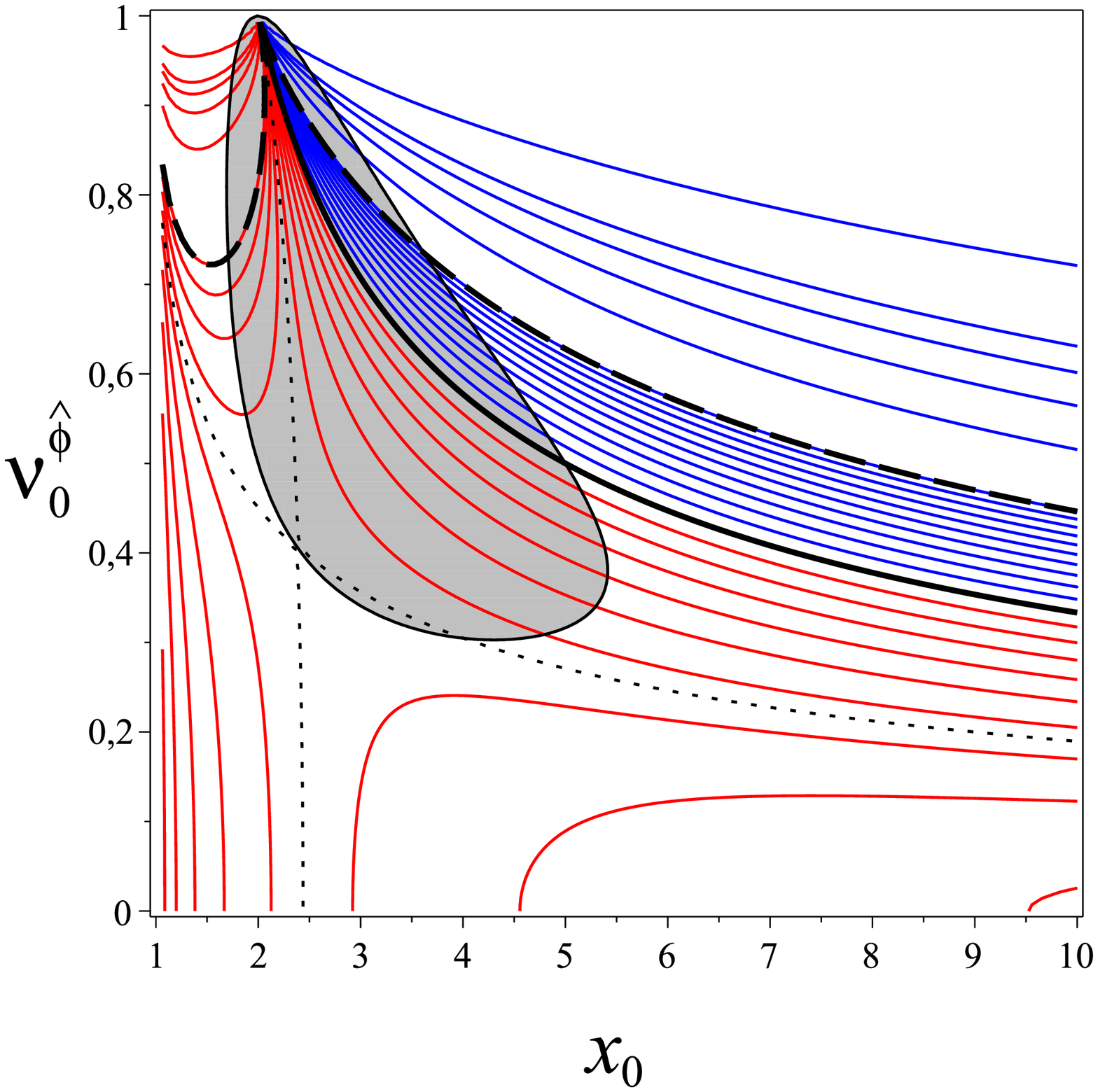}}
\hspace{5mm}
\subfigure{\includegraphics[scale=0.2]{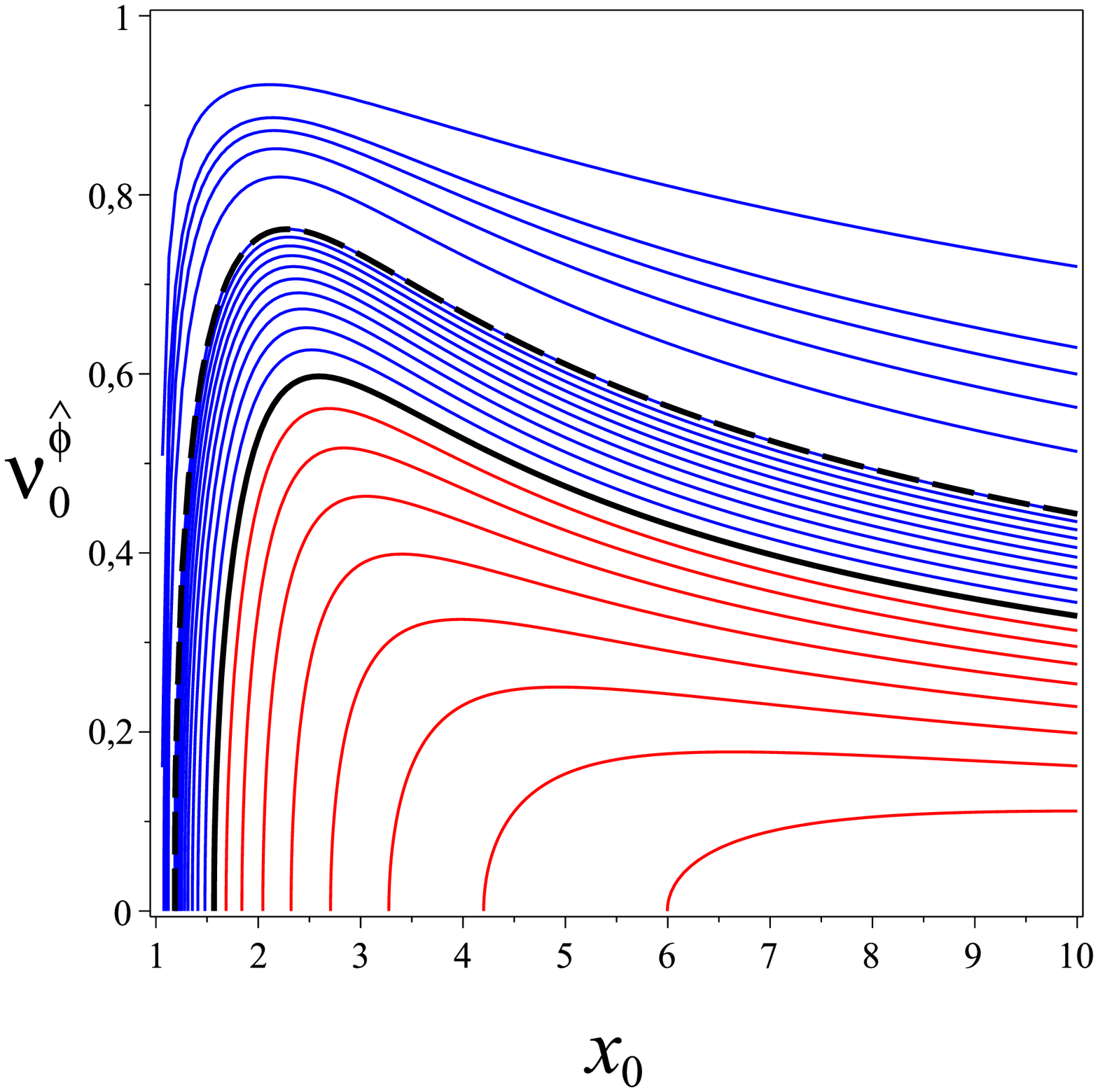}}
\hspace{5mm}
\subfigure{\includegraphics[scale=0.2]{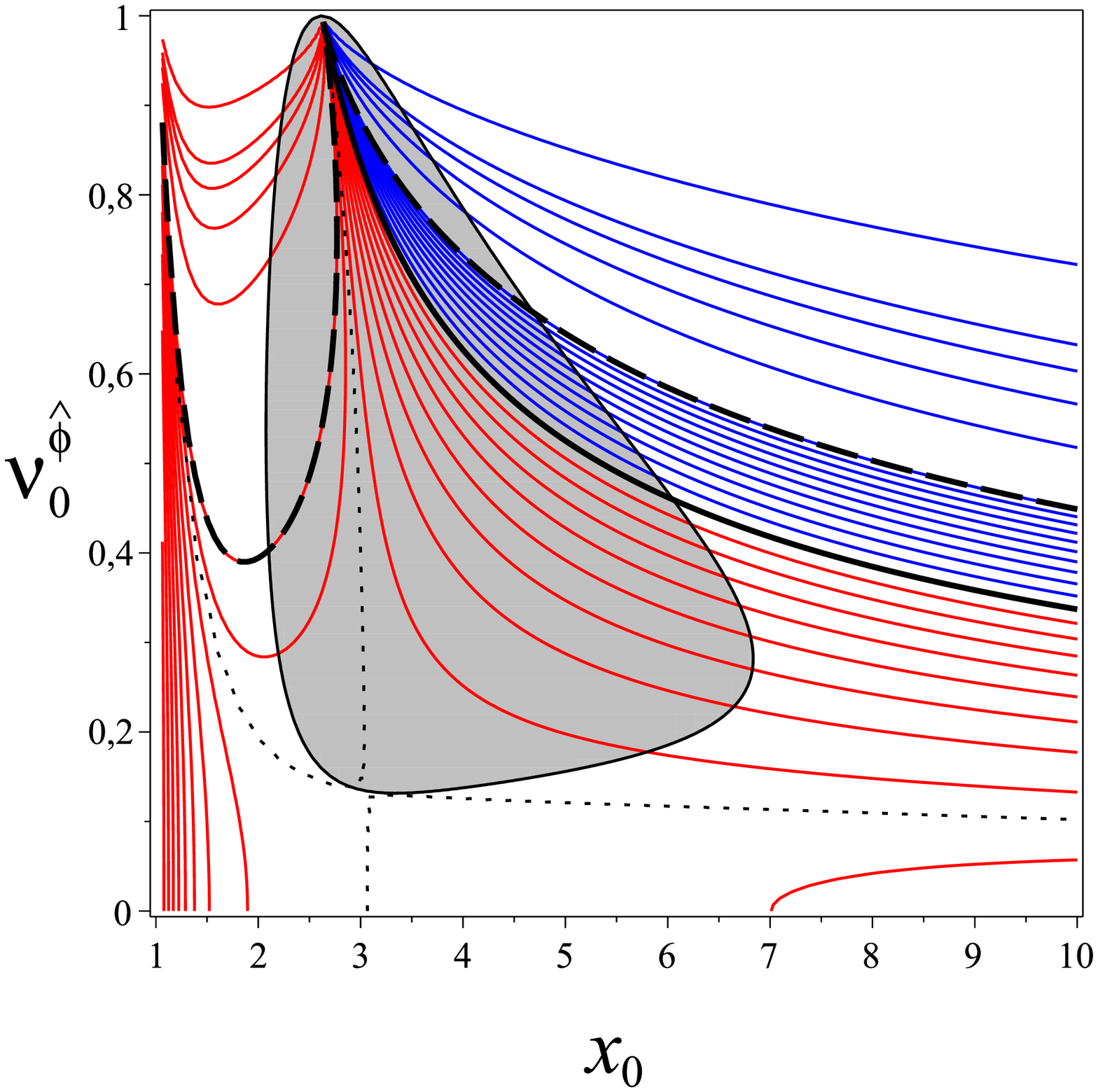}}\\[1cm]
\subfigure{\includegraphics[scale=0.2]{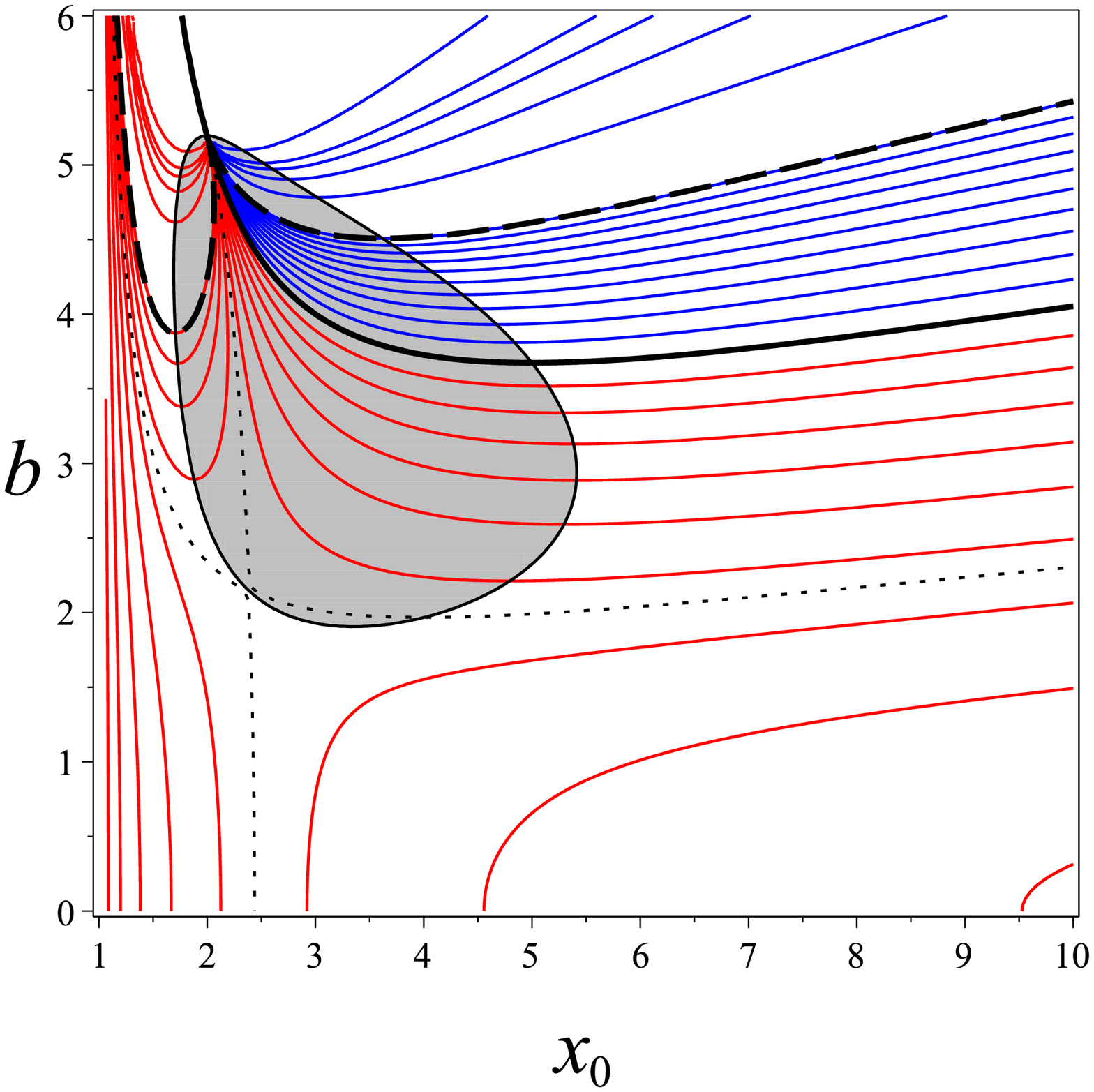}}
\hspace{5mm}
\subfigure{\includegraphics[scale=0.2]{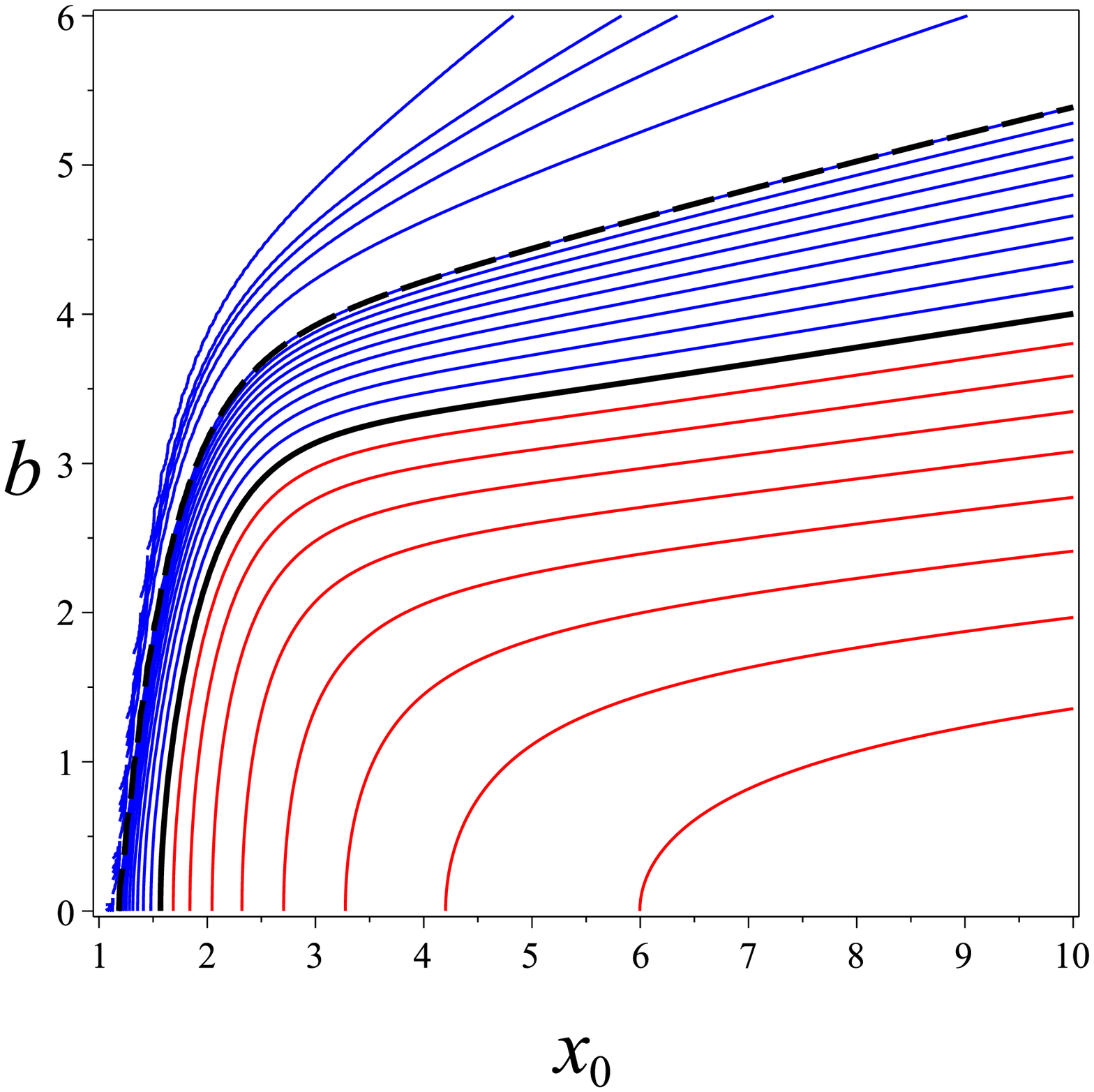}}
\hspace{5mm}
\subfigure{\includegraphics[scale=0.2]{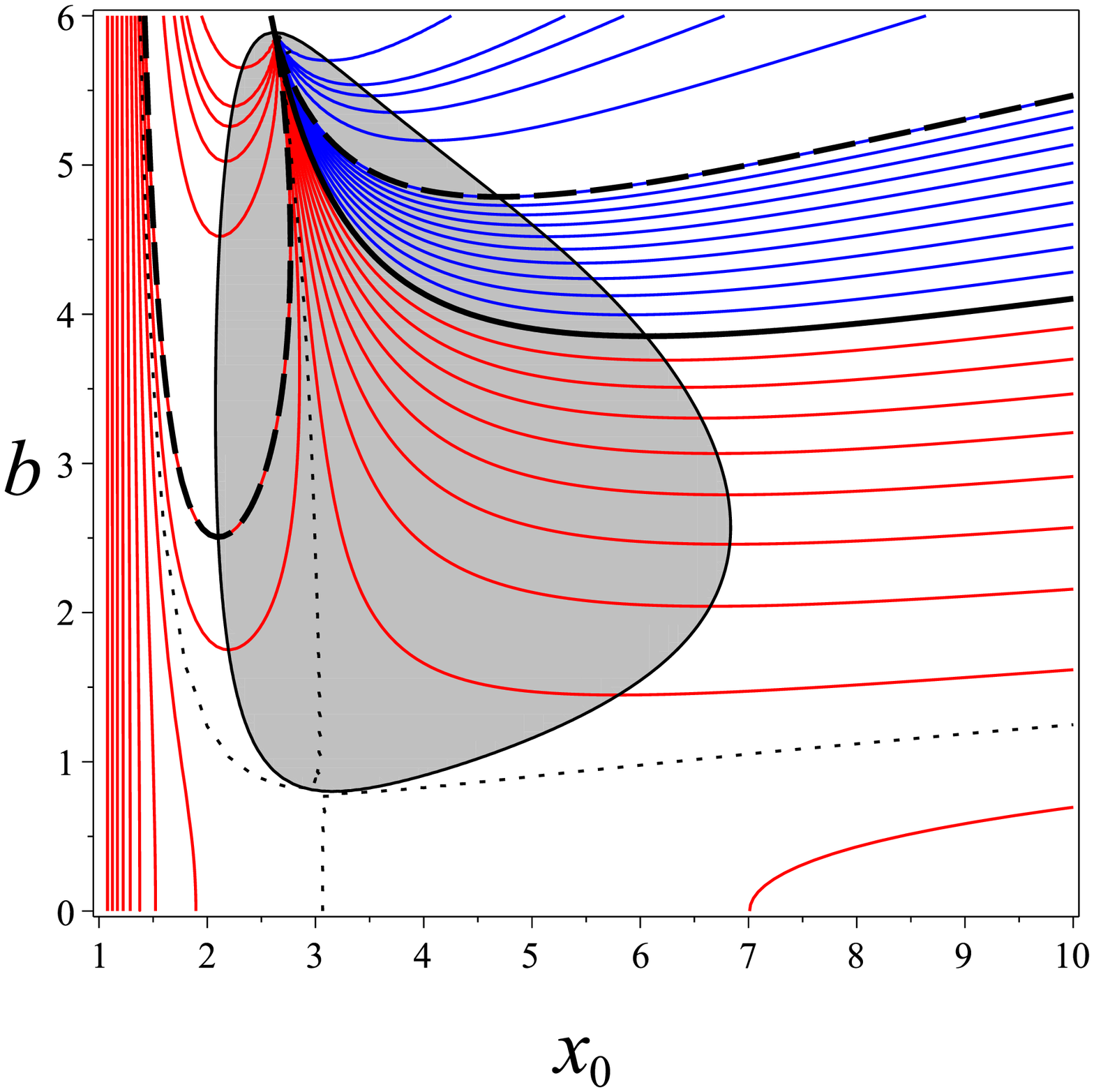}}
\caption{
{\bf First row}: Critical azimuthal velocity versus critical radius for equally spaced values of $A$ from 0 to 1 at intervals of $0.1$ and thereafter values of $[1, 2, 3, 4, 5, 10]$ (in units of $\sigma$) in both cases of outgoing flux ($\sin \beta_0>0$, red curves) and ingoing flux ($\sin \beta_0<0$, blue curves) and different values of the quadrupole parameter: $q=0$ (left), $q=10$ (middle) and $q=-10$ (right).
The thick solid curves indicate the geodesic velocities corresponding to $A = 0$, whereas the thick dashed curves correspond to $A = 1$.
The separatrix is for $A\approx0.647$ in the case $q=0$ and $A\approx0.850$ in the case $q=-10$. 
The thick loop curves enclose the region of unstable intermediate radius orbits for those values of $\nu^{\hat \phi}_0$ (or, equivalently, $b$) for which three critical radii exist.
{\bf Second row}: Critical photon impact parameter versus critical radius.  
}
\label{fig:equil}
\end{figure}

Fig. \ref{fig:equil} shows the behavior of the critical azimuthal velocity (first row) as well as critical photon impact parameter (second row) as a function of the radial distance for selected values of the friction parameter $A$ (in units of $\sigma$) and different values of the quadrupole parameter.
The critical speed is less/greater than the geodesic speed in the case of outgoing/ingoing photons ($\sin\beta_0>0\,/\,\sin\beta_0<0$).
The features of equilibrium are different depending on whether prolate or oblate sources are considered.
In the former case, for increasing values of $A$ the curves lie in the region above the geodesic velocity for ingoing photons, with the asymptotic limit $\nu^{\hat\phi}_0\to1$ for $A\to\infty$.
For outgoing photons, instead, the curves are below the geodesic one, continuing to move out far from the source for increasing values of $A$ without limit.
In the case of oblate sources, instead, the situation is very similar to the Schwarzschild one \citep{PR2}.
Consider first the case of outgoing photons.
The curves of constant $A$ start at the value $A=0$ at the hypersurface $x=1$ (vertical axis) and on the geodesic velocity curves, and increase in value towards the saddle point on the separatrix curve at a certain value of $A$. For $A$ further increasing past that value, the curves fill the inner sector above that point (near the hypersurface $x=1$).
At the same time, in the region below the meeting point they move farther out to larger radii, moving out to infinity in the limit $A\to1$, corresponding to the fact that for $A>1$, test particles are pushed out to infinity in this region.
In the case of ingoing photons, instead, all the curves of constant $A$ are confined in the region above the geodesic velocity.

The equilibrium condition (\ref{equil2}) can also be written in terms of the photon impact parameter in place of the critical speed, by using Eq. (\ref{nu0limit}).
The resulting equation determines the critical radius in terms of the photon impact parameter.
For fixed values of $b$ and $A$, one or more values of $x_0$ may satisfy it, which may correspond to stable orbits or not.
The stability analysis of critical orbits is done in the next section.   
For each curve in the plots shown in the second row of Fig.~\ref{fig:equil} corresponding to fixed values of $A$, the number of its intersections with the horizontal line at each value of $b$ indicates the number of critical radii which exist for that case. 
Those horizontal curves which pass through the interior of the closed shaded regions correspond to the case in which three critical radii exist, with the unstable orbit of the three lying in the interior, while those passing outside these loops correspond to the single stable critical radius case.

\begin{figure} 
\centering
\subfigure{\includegraphics[scale=0.2]{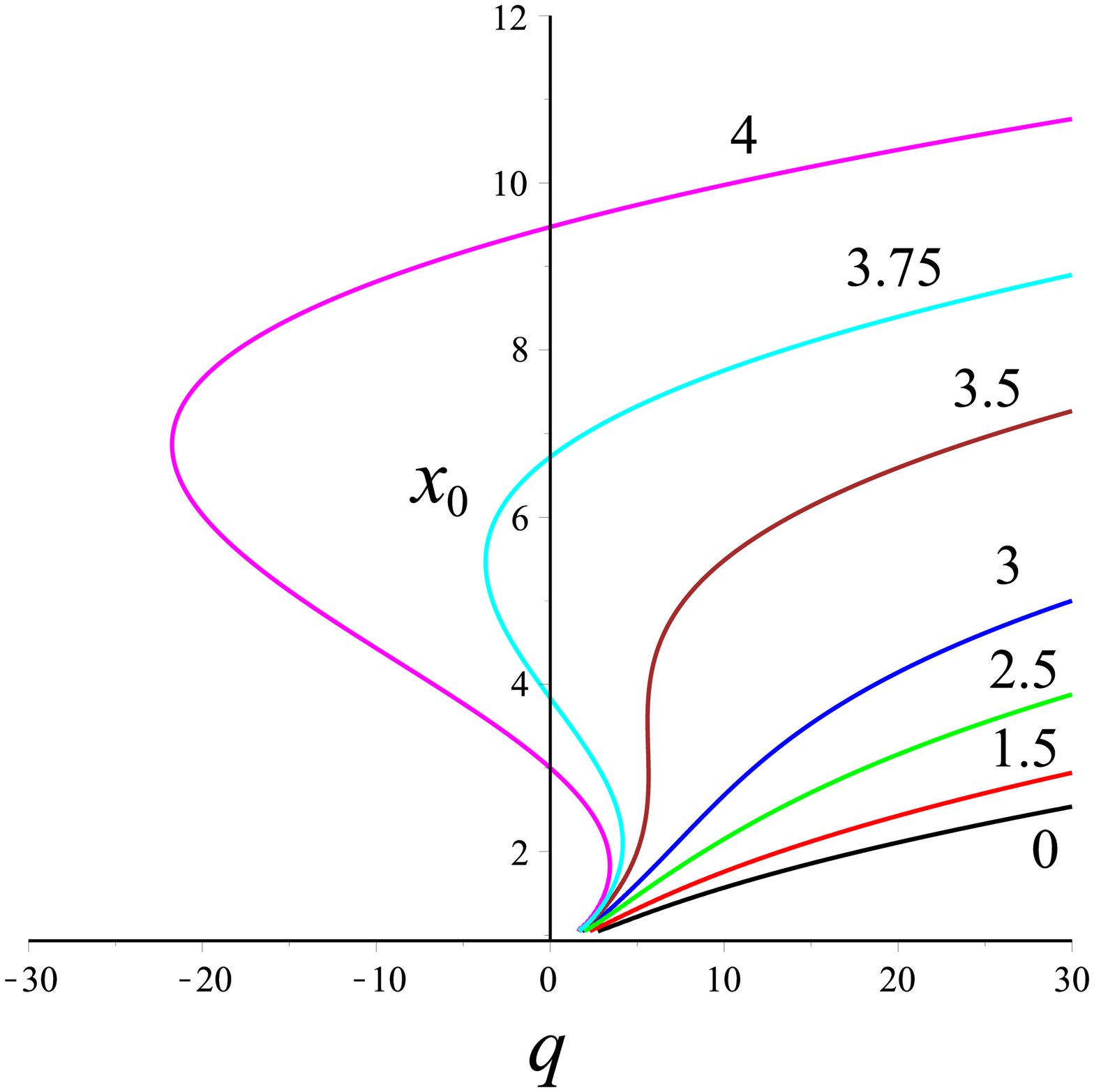}}
\hspace{5mm}
\subfigure{\includegraphics[scale=0.2]{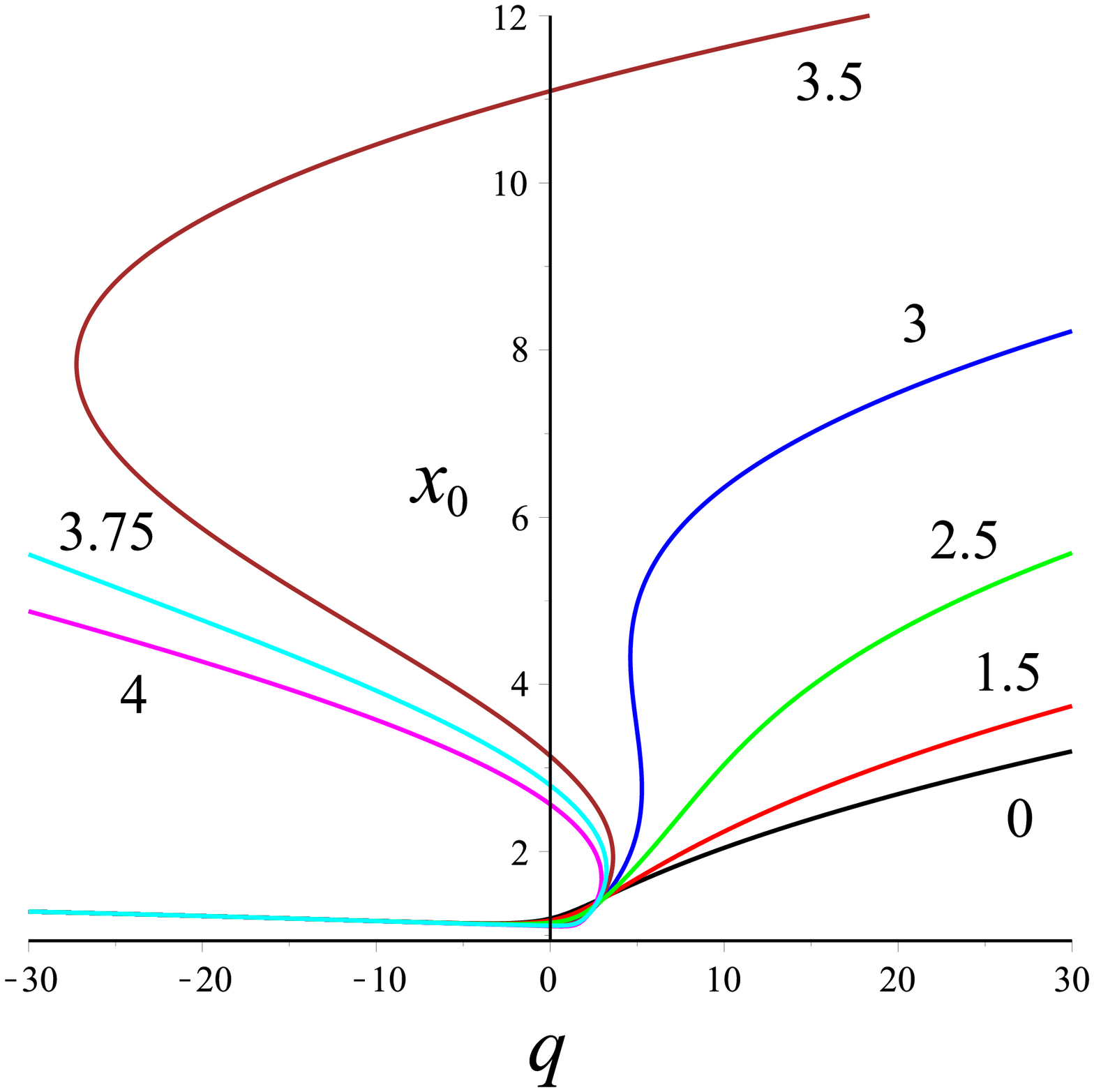}}
\hspace{5mm}
\subfigure{\includegraphics[scale=0.2]{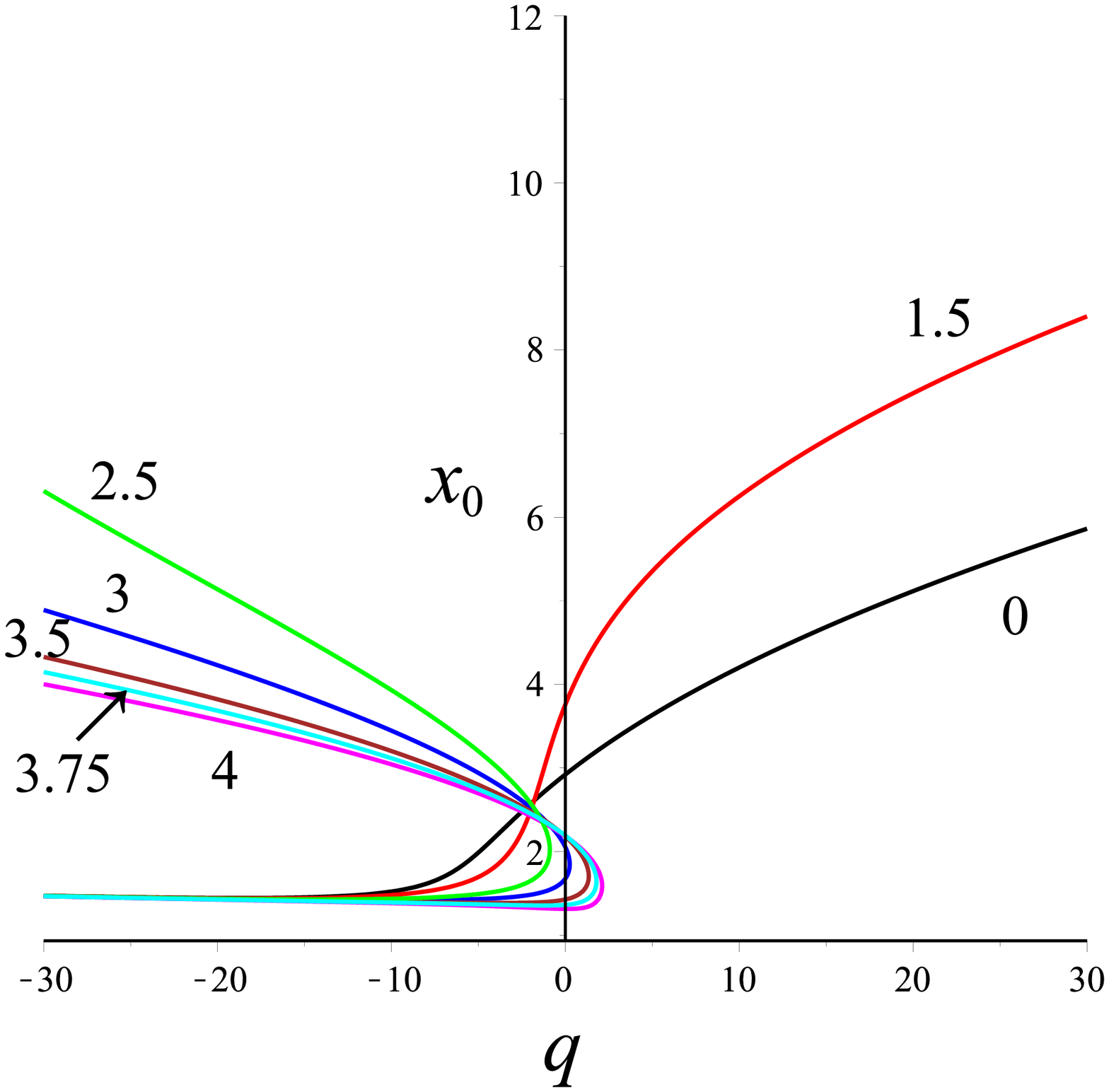}}
\caption{Critical radius versus quadrupole parameter for selected values of the strength of the radiation field $A = 0$ (left),  $A = 0.3$ (middle) and $A = 0.7$ (right) and different values of the photon impact parameter $b=[0,1.5,2.5,3,3.5,3.75,4]$. 
}
\label{fig:x_vs_q}
\end{figure}

Finally, Fig. \ref{fig:x_vs_q} shows the behavior of the critical radius as a function of the quadrupole parameter for selected values of the friction parameter $A$ and different values of the the photon impact parameter $b$ (in units of $\sigma$).
For small values of $b$ there exists in general a single equilibrium radius for every value of the quadrupole parameter, which is closer to the source as the configuration is more oblate.
For larger values of the photon impact parameter multiple solutions appear for $q\lesssim4$. 
Increasing the strength of the radiation field then causes the latter behavior to occur for even smaller values of $b$.


In the limit of vanishing photon impact parameter $b=0$ Eq. (\ref{nu0limit}) implies $\nu^{\hat \phi}_0=0$, which corresponds to \lq\lq suspension orbits,"i.e., particle trajectories end up at a certain point in the symmetry plane.
The equilibrium condition (\ref{equil}) reduces to
\beq
a(n)^{\hat x}=\frac{A}{\sigma^2}\frac{\Phi^2}{\Phi_0^2}\frac{1}{f}
=\frac{A}{\sigma^2}\frac{e^{-\gamma}}{x\sqrt{x^2-1}}
\,,
\eeq
and implicitly gives the suspension radius as a function of $A$ for fixed values of the quadrupole parameter.

\subsection{Stability of the critical orbits}

Let us study the stability properties of the critical circular orbits under small first order linear perturbations of the equilibrium solution, whose parametric equation is given by 
\beq
x=x_0\,, \quad 
\phi=\phi_0(\tau)\,, \quad
\nu^{\hat x}=0\,, \quad
\nu^{\hat \phi}=\nu^{\hat \phi}_0\,,
\eeq
or symbolically $X^\alpha=X_0^\alpha$ ($\alpha=x,\phi,\nu^{\hat x},\nu^{\hat \phi}$). 
Note that this analysis will not check for stability against perturbations away from the equatorial plane.

Consider the linear perturbations of this solution $X^\alpha=X_0^\alpha+X_1^\alpha$, namely
\beq
x=x_0+x_1(\tau)\ , \quad 
\phi=\phi_0(\tau)+\phi_1(\tau), \quad
\nu^{\hat x}=\nu^{\hat x}_1(\tau)\ , \quad
\nu^{\hat \phi}=\nu^{\hat \phi}_0+\nu^{\hat \phi}_1(\tau)\,,
\eeq
which leads to the following linear system of constant coefficient homogeneous linear differential equations
\beq
\frac{\rmd X_1^\alpha}{\rmd\tau}=C^\alpha{}_\beta X_1^\beta\,,
\eeq
which can easily be solved in terms of the eigenvalues and eigenvectors of the coefficient matrix. The real parts of all eigenvalues must be nonnegative for stability.
The explicit expressions for the nonzero coefficients are
\begin{eqnarray}
C^1{}_{3}&=&\frac{\gamma_0}{\sqrt{g_{xx}}}\,,\qquad
C^2{}_{1} = \gamma_0\nu^{\hat \phi}_0k(\phi,n)^{\hat x}\frac{\sqrt{g_{xx}}}{\sqrt{g_{\phi\phi}}}\,,\qquad
C^2{}_{4} = \frac{\gamma_0^3}{\sqrt{g_{\phi\phi}}}
\,,\nonumber\\
C^3{}_{1}&=& -\gamma_0\sqrt{g_{xx}}\left\{
\left(E(n)_{\hat x\hat x}-a(n)^{\hat x}{}^2\right)-\nu^{\hat \phi}_0{}^2\left(E(n)_{\hat y\hat y}-k(\phi,n)^{\hat x}{}^2\right)\right.\nonumber\\
&&\left.
+C^4{}_{4}\left[-(2-\nu^{\hat \phi}_0{}^2)a(n)^{\hat x}+(1-2\nu^{\hat \phi}_0{}^2)k(\phi,n)^{\hat x}+(1-\nu^{\hat \phi}_0{}^2)k(y,n)^{\hat x}\right]
\right\}
\,,\nonumber\\
C^3{}_{3} &=& -2C^4{}_{4}\,,\qquad
C^3{}_{4}=-2\gamma_0^3\nu^{\hat \phi}_0(a(n)^{\hat x}+k(\phi,n)^{\hat x})
\,, \nonumber\\
C^4{}_{1}&=&\frac{\sqrt{g_{xx}}}{2\gamma_0^3}C^3{}_{4}C^4{}_{4}\,,\qquad
C^4{}_{3}=\frac{\nu^{\hat \phi}_0}{\gamma_0}(k(\phi,n)^{\hat x}-C^4{}_{4})\,,\qquad
C^4{}_{4} = \frac{A\sigma^2}{\gamma_0f\sqrt{g_{yy}}\sqrt{g_{\phi\phi}}}\,,
\end{eqnarray} 
with $A$ given by Eq. (\ref{equil2}).
The quantities $E(n)_{\hat a\hat b}$ in the above equations are the frame components with respect to $n$ of the electric part of the Riemann tensor defined by $E(n)_{\alpha\beta}=R_{\alpha\mu\beta\nu}n^\mu n^\nu$.

The associated eigenvalue equation is 
\beq
\label{eqlambda}
\lambda[\lambda^3+c_2\lambda^2+c_1\lambda+c_0]=0\,,
\eeq
where
\begin{eqnarray}
\frac{c_0}{c_2}+c_1+2c_2^2&=&2\frac{\gamma_0^2}{\gamma_{\rm g}^2}\frac{\nu^{\hat \phi}_0{}^2}{\nu_{\rm g}^2}a(n)^{\hat x}\left[
c_2-\gamma_0^2(\nu^{\hat \phi}_0{}^2-\nu_{\rm g}^2)\frac{a(n)^{\hat x}}{\nu_{\rm g}^2}
\right]
\,,\nonumber\\
c_1+2c_2^2 &=&\frac{\gamma_0^2}{\gamma_{\rm g}^2}\Omega_{\rm (ep)}^2
+\gamma_0^2(\nu^{\hat \phi}_0{}^2-\nu_{\rm g}^2)\left[
-E(n)_{\hat y\hat y}+\frac{a(n)^{\hat x}{}^2}{\nu_{\rm g}^4}(3-2\nu_{\rm g}^2)
\right]\nonumber\\
&&
-c_2\left[k(y,n)^{\hat x}-\gamma_0^2a(n)^{\hat x}\left(2-3\nu^{\hat \phi}_0{}^2+\frac1{\nu_{\rm g}^2}\right)\right]
\,, \nonumber\\
c_2&=&C^4{}_{4}
\,,
\end{eqnarray}
and 
\beq
\label{eq:Omegaep}
\Omega_{\rm (ep)}^2=3\frac{a(n)^{\hat x}{}^2}{\nu_{\rm g}^2}+\gamma_{\rm g}^2(E(n)_{\hat x\hat x}-\nu_{\rm g}^2E(n)_{\hat y\hat y})\,, 
\eeq
is the corresponding proper time normalized version of the time coordinate epicyclic frequency governing the radial perturbations of circular geodesics.
In the Schwarzschild case it reduces to 
\beq
\Omega_{\rm (ep)}^2=\frac{x_0-5}{\sigma^2(x_0-2)(x_0+1)^3}
=\frac{M}{r_0^3}\frac{r_0-6M}{r_0-3M}\,.
\eeq
The explicit expressions for the eigenvalues can be obtained straightforwardly.
In the case of geodesic motion (i.e., $A=0$ and $\nu^{\hat \phi}_0=\pm\nu_{\rm g}$) the eigenvalue equation (\ref{eqlambda}) reduces to 
\beq
\lambda^2(\lambda^2+\Omega_{\rm (ep)}^2)=0\,,
\eeq
since $c_0=0=c_2$ and $c_1=\Omega_{\rm (ep)}^2$.

\begin{figure}
\centering
\subfigure{\includegraphics[scale=0.25]{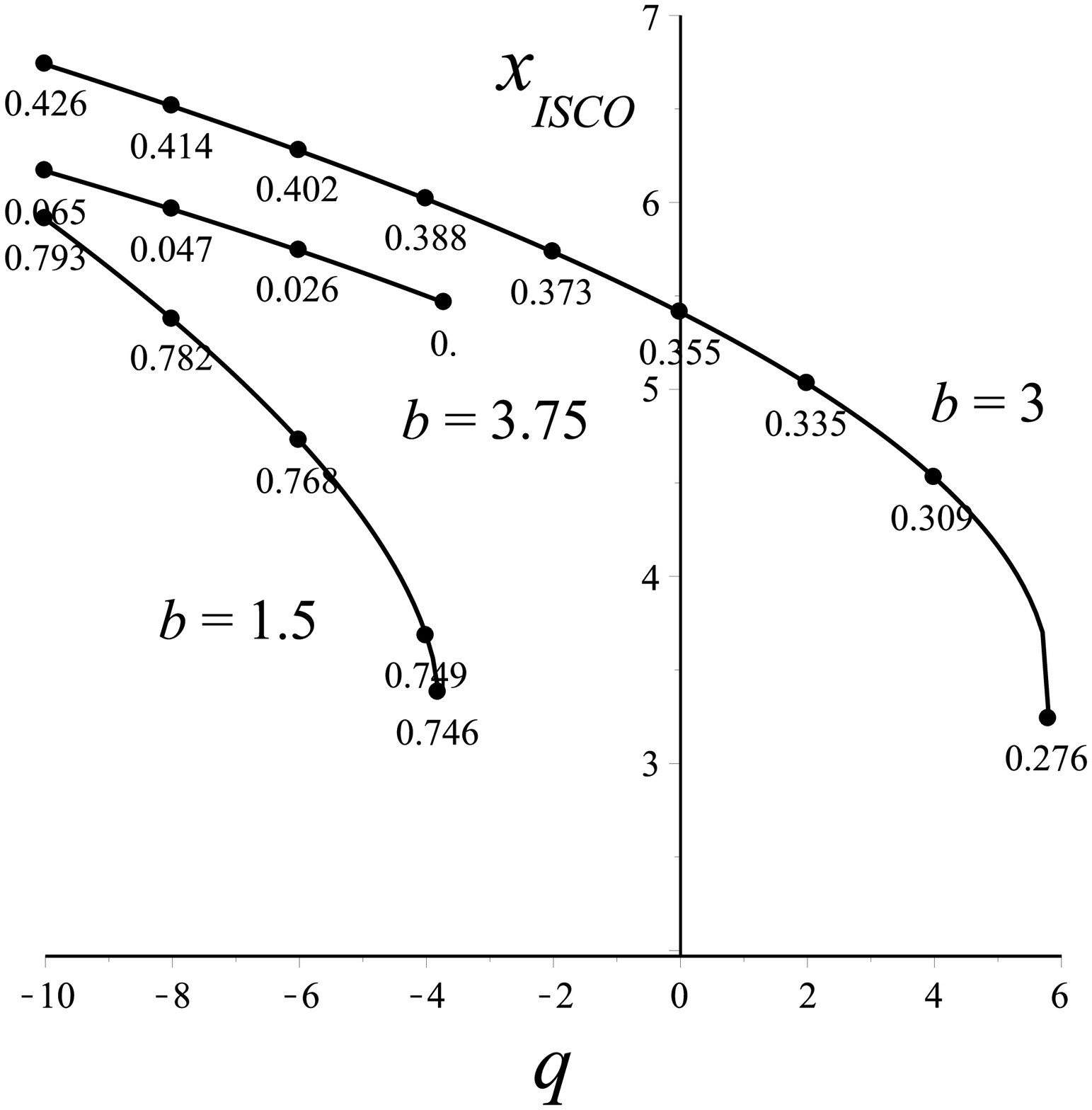}}
\hspace{5mm}
\subfigure{\includegraphics[scale=0.25]{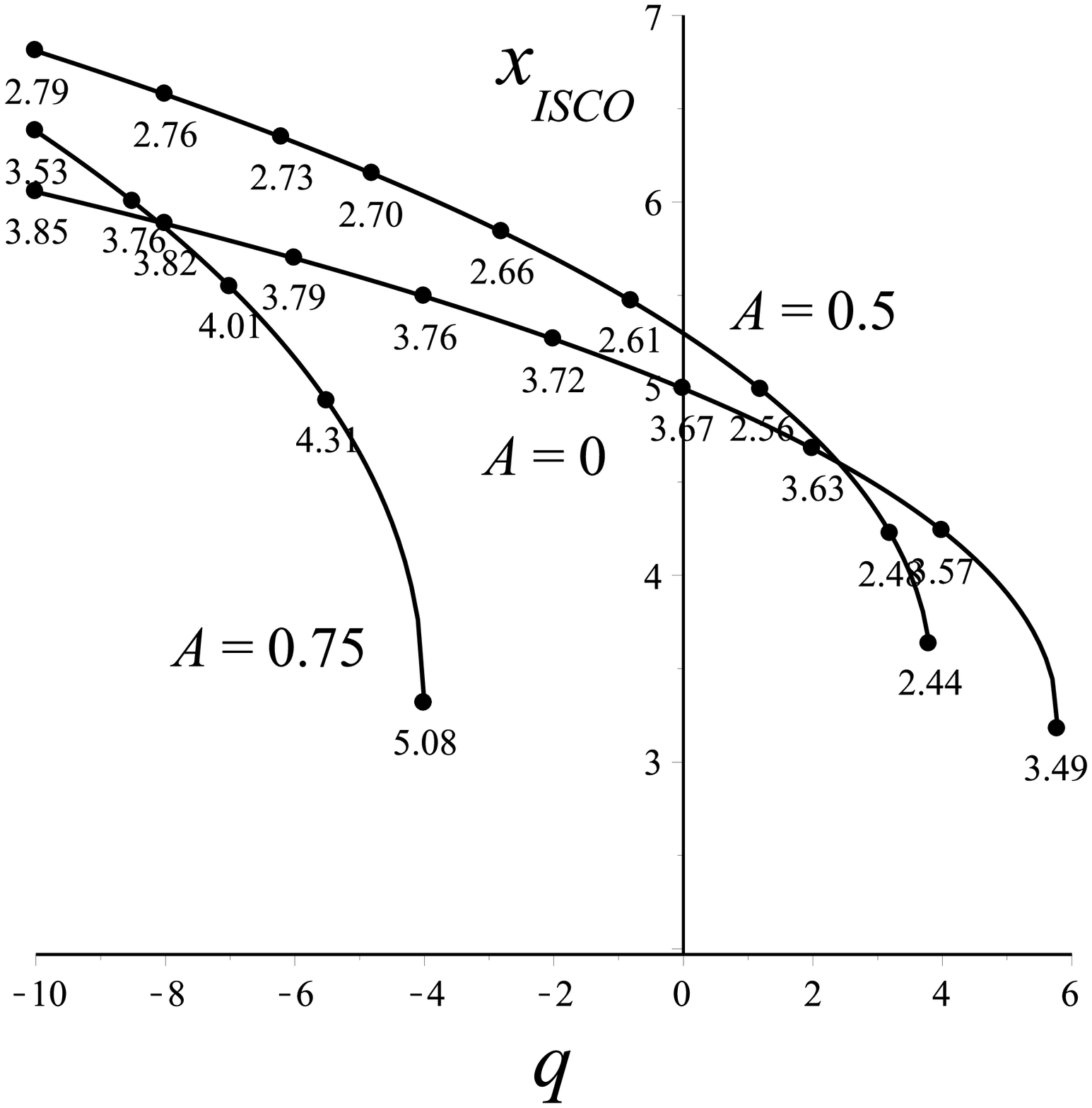}}
\caption{The behavior of the ISCO radius as a function of the quadrupole parameter is shown for outgoing photon flux for fixed values of $b=[1.5,3,3.75]$ in the left panel and for fixed values of $A=[0,0.5,0.75]$ in the right panel (in units of $\sigma$).
Each point in the curves corresponds to a given value of $A$ in the former case, and to a given value of $b$ in the latter case.
A selection of points identified by a dot with the value of either $A$ or $b$ indicated below is also shown.
}
\label{fig:isco}
\end{figure}

The behavior of the eigenvalues as functions of the radial distance from the central source shows that for a given value of the quadrupole parameter there exist in general two disconnected regions where circular orbits are stable (see Fig. \ref{fig:equil}): one very close to the singularity, and another for larger radii, which is relevant for observational effects.
Therefore, we refer to the left boundary of that region as the radius corresponding to the innermost stable orbit (ISCO).
The behavior of the ISCO radius as a function of the quadrupole parameter is shown in Fig. \ref{fig:isco} for both fixed values of $b$ (left panel) and fixed values of $A$ (right panel). 
We find that the orbits are always stable for prolate configurations with $q\gtrsim6$.


\section{Concluding remarks}

In any attempt to construct a realistic model of test particle motion around compact objects to be constrained by observations one should take into account both the structure of these objects and the effects of the presence of electromagnetic or radiation fields surrounding them.
Black holes are commonly expected to be hosted at the center of galaxies as well as in X-ray binary systems (see, e.g., \citet{narayan} and references therein).
According to the Einstein's theory of gravity, an uncharged black hole is completely specified by its total mass and angular momentum, which determine all the higher order mass and current moments of the gravitational field.
In particular, the mass quadrupole moment of a spinning black hole is proportional to the angular momentum squared.
However, general relativity allows for exact solutions which represent compact objects with additional structure, i.e., with the multipole moments of the associated gravitational field expressed in terms of an arbitrary set of parameters \citep{ES}.
For instance, the Erez-Rosen solution, which we adopt in the present paper, generalizes the Schwarzschild spacetime to the case of a gravitational source endowed with an arbitrary mass quadrupole moment, and hence it is specified by two parameters, the mass $M$ and the quadrupole parameter $q$.
As expected, the presence of the quadrupole affects the motion of test particles and photons in a significant way \citep{armenti,def,quev90,ERlight}.

There is a growing interest in the current literature in constraining the geometry around compact objects with observations in both the electromagnetic and gravitational wave spectra.
In fact, both space and ground-based advanced gravitational wave detectors (like LISA \citep{lisa} and advanced LIGO/Virgo \citep{ligo,virgo}) are expected to detect, e.g., the gravitational waves emitted during the inspiral of binary systems.
Any deviation from either the Schwarzschild or Kerr black hole geometry will then lead to a phase difference in the gravitational waveforms as well as to the  \lq\lq ringdown radiation'' in the form of the so-called quasinormal modes of the end state of the merger \citep{ryan,barack,rodriguez,berti}.

It is also a general feature that compact objects are surrounded by an accretion disc, which forms due to the diffuse material orbiting them coming from either the companion in the case of stellar-mass objects in X-ray binary systems, or from the interstellar medium in the case of supermassive objects in galactic nuclei.
For these systems the effect of a nonvanishing quadrupolar structure can also be inferred from available X-ray data, by studying for instance relativistic broadening of iron lines \citep{iron}, disc's thermal spectrum \citep{dspec}, quasi-periodic oscillations \citep{qpos}, X-ray polarization \citep{xpola} and mean radiative efficiency of AGN \citep{eta}.
In many of the existing models, accretion flows are represented by thin discs lying in the equatorial plane of the compact object, and the particles in the disc typically move on circular orbits.
Since the spacetime curvature experienced by the disc is largest at the inner edge, where most of the observed flux is emitted, its location is of special importance.
It is usually taken to coincide with the innermost stable circular orbit (ISCO), which depends on the mass and angular momentum in the case of a Kerr black hole, but may depends on other spacetime parameters, if more general solutions are considered.

In the present paper, we have studied the motion of test particles undergoing Poynting-Robertson effect (i.e., scattered by a superposed test radiation field) in the field of a static non-spherical source described by the Erez-Rosen solution.
The radiation field is modeled by photons which move on the equatorial plane and have an arbitrary, but constant angular momentum.
The particle-photon interaction produces non-geodesic orbits around the central body whose features depend both on the radiation field strength and the photon impact parameter, but also on the value of the mass quadrupole parameter of the source. 
In fact, particles as well as photons are expected to feel a gravitational field with different strength in the region close to the source with respect to the spherically symmetric case ($q=0$) depending on whether the central object is oblate ($q<0$) or prolate ($q>0$).
In the former case the incoming particle feels a stronger gravitational field as it approaches the gravity source.
For prolate configurations, instead, the gravitational mass is mainly concentrated along the $y-$axis, i.e., orthogonally to the symmetry plane where the motion takes place, so that a particle moving sufficiently close to the gravity source will feel a decreasing gravitational field, since most of the gravitational action is being neutralized there.

The main outcome of our analysis is that there exists a whole family of equilibrium solutions (circular orbits) representing the balance between gravitational attraction, centrifugal force and radiation drag.
In fact, if the strength of the radiation field is fixed, the quadrupole parameter of the source can be used to parametrize this family, so leading to a scenario which is more rich with respect to that of a spherically symmetric source.
The associated stability analysis has allowed to determine the ISCO radius as a function of the parameters characterizing the radiation field as well as of the quadrupole parameter of the source.  
We have found that for a given value of the quadrupole parameter there exist in general two disconnected regions where circular orbits are stable: one very close to the central object, and another for larger radii, which is relevant for observational effects. 
This analysis can be straightforwardly extended to more general situations, e.g., by accounting for quadrupolar sources also endowed with nonvanishing rotation.

\appendix

\section{Geometrical properties of the Erez-Rosen solution}

The Legendre polynomials entering the definition of the metric functions (\ref{metdef}) are listed below:
\begin{eqnarray}
P_0(z)&=&1\,,\qquad
P_1(z)=z\,,\qquad
P_2(z)=-\frac12(1-3z^2)\,,\nonumber\\
Q_0(z)&=&\frac12\ln\left(\frac{z+1}{z-1}\right)\,,\qquad
Q_1(z)=zQ_0(z)-1\,,\qquad
Q_2(z)=-\frac12[Q_0(z)-3zQ_1(z)]\,.
\end{eqnarray}

According to the Geroch-Hansen \citep{ger,hans} definition of relativistic multipole moments ${\mathcal M}_n$, the mass monopole moment associated with this solution is ${\mathcal M}_0=M$ and the quadrupole moment is given by ${\mathcal M}_2=(2/15)qM^3$.
Higher multipole moments of the order $n=2k$, $k=2,3,4,\ldots$, are determined by $q$ and $M$ in such a way that they all vanish when $q=0$.

The presence of the quadrupole parameter $q$ changes significantly the structure of the spacetime as compared with the Schwarzschild solution. In particular the hypersurface $x=1$, which is null in the Schwarzschild case, becomes directionally singular.
Furthermore, its character depends both on the value of $q$.
For instance, in the symmetry plane $y=0$ this hypersurface is null for $1-\sqrt{5}<q<1+\sqrt{5}$ and timelike otherwise (see, e.g., \citep{quev90,masquev95}).

The explicit expressions for the radial components of the acceleration as well as curvature vectors in the symmetry plane $y=0$ are listed below:
\begin{eqnarray}
a(n)^{\hat x}
&=&\frac12\frac{\sqrt{x^2-1}}{\sigma x}e^{-\gamma}\frac{f_x}{\sqrt{f}}\,, \nonumber \\
k(x,n)^{\hat x}
&=&k(\phi,n)^{\hat x}-e^{-\gamma}\frac{\sqrt{f}}{\sigma x^2\sqrt{x^2-1}}[(x^2-1)(x\gamma_x-1)-2]\,, \nonumber \\
k(y,n)^{\hat x}
&=&k(\phi,n)^{\hat x}-e^{-\gamma}\frac{\sqrt{f}}{\sigma x^2\sqrt{x^2-1}}[(x^2-1)x\gamma_x-1]\,, \nonumber \\
k(\phi,n)^{\hat x}
&=&a(n)^{\hat x}-e^{-\gamma}\frac{\sqrt{f}}{\sigma\sqrt{x^2-1}}\,,
\end{eqnarray}
with 
\begin{eqnarray}
f&=&\frac{x-1}{x+1}e^{qQ_2}\,,\qquad
\frac{f_x}{f}=\frac{2}{x^2-1}+qQ_2'\,,\qquad
\gamma_x=\frac{(x^2-1)f_x^2}{4xf^2}\,, \nonumber\\
\frac{f_{xx}}{f}&=&-\frac{4}{(x^2-1)(x+1)}+q\left(Q_2''+\frac{4}{x^2-1}Q_2'\right)+q^2Q_2'{}^2\,, \qquad
\frac{f_{yy}}{f}=-6qQ_2\,,
\end{eqnarray}
and $Q_2$ satisfies the following second order differential equation
\beq
Q_2''+\frac{2x}{x^2-1}Q_2'-\frac{6}{x^2-1}Q_2=0\,.
\eeq

Finally, the relevant frame components of the electric part of the Riemann tensor are given by
\begin{eqnarray}
E(n)_{\hat x\hat x}
&=&\frac12\frac{e^{-2\gamma}}{\sigma^2x^2}\left[(x^2-1)(f_{xx}-\gamma_xf_x)+\frac{f_x}{f}\right]\,, \nonumber \\
E(n)_{\hat y\hat y}
&=&-E(n)_{\hat x\hat x}+\frac14\frac{e^{-2\gamma}}{\sigma^2x^2}\frac{f_x}{f}\left[(x^2-1)f_x-2xf\right]\,.
\end{eqnarray}

\section*{Acknowledgements}

DB and AG acknowledge ICRANet for partial support.
All the authors thank Prof. L. Stella for useful discussions at the beginning of the present project.
AP acknowledges support from the European Union Seventh Framework Programme (FP7/2007-2013) under grant agreement no. 267251 'Astronomy Fellowships in Italy (AstroFIt).'

\label{lastpage}

\end{document}